\documentclass[12pt]{article}
\usepackage{color}
\usepackage{amssymb,amsmath,comment}
\usepackage{graphicx}
\usepackage{braket}
\usepackage{epsfig}
\usepackage{here}
\usepackage{bm}
\graphicspath{{./Figures/}}
\newcommand{\ba}{\begin{align}}
\newcommand{\ea}{\end{align}}

\def\ov{\overline}
\def\nn{\nonumber}

\def\bea{\begin{eqnarray}}
\def\eea{\end{eqnarray}}
 \makeatletter
\def\alt{\mathrel{\mathpalette\gl@align<}}
\def\agt{\mathrel{\mathpalette\gl@align>}}
\def\gl@align#1#2{\lower.6ex\vbox{\baselineskip\z@skip\lineskip\z@
\ialign{$\m@th#1\hfil##\hfil$\crcr#2\crcr\sim\crcr}}} \makeatother
\renewcommand{\thefootnote}{\fnsymbol{footnote}}
\begin{document}
\begin{flushright}
\end{flushright}
\vspace*{1.0cm}

\begin{center}
\baselineskip 20pt 
{\Large\bf 
Neutrino Mass in Non-Supersymmetric $SO(10)$ GUT
}
\vspace{1cm}

{\large 
Naoyuki Haba$^a$, Yasuhiro Shimizu$^a$ and Toshifumi Yamada$^b$
} \vspace{.5cm}

{\baselineskip 20pt \it
$^a$ Department of Physics, Osaka Metropolitan University, Osaka 558-8585, Japan\\
$^b$ Department of Physics, Yokohama National University, Yokohama 240-8501, Japan
}

\vspace{.5cm}

\vspace{1.5cm} {\bf Abstract} \end{center}

We study a prediction on neutrino observables in a non-supersymmetric renormalizable $SO(10)$ GUT model
 that contains a {\bf 10} complex scalar field and a {\bf 126} scalar field whose Yukawa couplings with {\bf 16} matter fields
 provide the quark and charged lepton Yukawa couplings, neutrino Dirac Yukawa coupling and Majorana mass for the singlet neutrinos.
The $SO(10)$ breaking is achieved in two steps by a ${\cal O}(10^{15})$~GeV VEV of a {\bf 54} real scalar field and a ${\cal O}(10^{14})$~GeV VEV of the {\bf 126} field.
First, we analyze the gauge coupling unification conditions and determine the VEV of the {\bf 126} field.
Next, we constrain the Yukawa couplings of the ${\bf 10}$ and ${\bf 126}$ fields at the scale of the {\bf 126} field's VEV
 from experimental data on quark and charged lepton masses and quark flavor mixings.
Then we express the active neutrino mass with the above Yukawa couplings and the {\bf 126} field's VEV based on the Type-1 seesaw mechanism,
 and fit neutrino oscillation data, thereby deriving a prediction on poorly or not measured neutrino observables.
What distinguishes our work from previous studies is that we do not assign Peccei-Quinn charges on visible sector fields 
 so that the {\bf 10} scalar field and its complex conjugate both have Yukawa couplings with {\bf 16} matter fields.
From the fitting of neutrino oscillation data, we find that not only the normal neutrino mass hierarchy, but also the inverted hierarchy can be realized.
We also reveal that in the normal hierarchy case, the Dirac CP phase of the neutrino mixing matrix $\delta_{CP}$ is likely in the ranges of
 $-2.4<\delta_{\rm CP}<-1.2$ and $1.2<\delta_{\rm CP}<2.4$, and not in the region with $\delta_{\rm CP}\sim\pi$,
 and that in the normal hierarchy case, $\theta_{23}$ is likely in the upper octant and in the range of $0.50\lesssim\sin^2\theta_{23}\lesssim0.55$.

\thispagestyle{empty}

\newpage
\renewcommand{\thefootnote}{\arabic{footnote}}
\setcounter{footnote}{0}
\baselineskip 18pt

\section{Introduction}

The $SO(10)$ grand unified theory (GUT)~\cite{Georgi:1974my,Fritzsch:1974nn} is a viable candidate for physics beyond the Standard Model (SM),
 because it accounts for the origin of the hypercharge quantization, and it is automatically equipped with the seesaw mechanism that naturally explains the tiny neutrino mass~\cite{Yanagida:1979as,Yanagida:1979gs,Gell-Mann:1979vob,seesaw4}.
There are four classes of $SO(10)$ GUT models which are either supersymmetric or non-supersymmetric and either renormalizable or non-renormalizable.
In supersymmetric models, the gauge coupling unification is achieved without intermediate scale, 
 whereas in non-supersymmetric models, one or more intermediate scales are necessary for the successful unification.
In renormalizable models~\cite{Babu:1992ia,Lee:1994je}, one introduces ${\bf 10}$ and ${\bf 126}+{\bf \ov{126}}$ (and optionally ${\bf 120}$) representation fields from which the electroweak-symmetry-breaking Higgs field originates, 
 and the renormalizable couplings of ${\bf 10}$ and ${\bf \ov{126}}$ with ${\bf 16}$ matter fields give rise to realistic SM Yukawa couplings
 and neutrino Dirac Yukawa coupling.
Additionally, the renormalizable coupling of ${\bf \ov{126}}$ and its vacuum expectation value (VEV) generate Majorana mass for the singlet neutrinos.
In non-renormalizable models, one introduces ${\bf 10}$ and ${\bf 16}+{\bf \ov{16}}$ fields (let us denote the latter by ${\bf 16}_H+{\bf \ov{16}}_H$ to avoid confusion),
 and the renormalizable coupling of ${\bf 10}$ with ${\bf 16}$ matter fields
 and the non-renormalizable couplings of two ${\bf \ov{16}}_H$'s with ${\bf 16}$ matter fields, combined with the VEV of ${\bf \ov{16}}_H$, generate realistic SM Yukawa couplings, neutrino Dirac Yukawa coupling and Majorana mass for the singlet neutrinos.
The renormalizable models are attractive 
 because the flavor structures of the neutrino Dirac Yukawa coupling and Majorana mass term
 can be constrained from experimental data on quark and charged lepton masses and quark flavor mixings,
 so that one can make a restrictive prediction on the neutrino mass and mixings.
However, the supersymmetric renormalizable models confront a serious trouble that the $SO(10)$ gauge coupling becomes non-perturbative near the unification scale because the Dynkin index of ${\bf 126}$ representation is large and both scalar and fermionic
 components contribute to the renormalization group (RG) evolution of the gauge coupling.
For the above reasons, the non-supersymmetric renormalizable models are worth for scrutiny.

In this paper, we study a non-supersymmetric renormalizable $SO(10)$ GUT model
 with emphasis on its prediction on the neutrino mass and mixings.
Specifically, we consider a non-supersymmetric $SO(10)$ GUT model containing a ${\bf 10}$ complex scalar field and a ${\bf 126}$ scalar field that couple with {\bf 16} matter fields.
The GUT breaking is achieved by a high-scale (${\cal O}(10^{15})$~GeV) VEV of a ${\bf 54}$ real scalar field and an intermediate-scale (${\cal O}(10^{14})$~GeV)
 VEV of the ${\bf 126}$ field~\cite{Babu:2015bna}.
Our analysis proceeds as follows: First, we analyze the gauge coupling unification conditions and evaluate the intermediate-scale VEV.
Next, we constrain the Yukawa couplings of the ${\bf 10}$ and ${\bf 126}^*$ fields at the intermediate scale from experimental data on the quark and charged lepton masses and quark flavor mixings.
Then we express the neutrino Dirac Yukawa coupling and Majorana mass for the singlet neutrinos with these Yukawa couplings and the intermediate-scale VEV 
 and calculate the neutrino mass matrix based on the Type-1 seesaw model.
Finally, we fit experimental data on the neutrino mixing angles and mass differences,
 and make a prediction on poorly or not measured neutrino observables.

Previously, non-supersymmetric renormalizable $SO(10)$ GUT models and their implication on neutrino physics have been studied in a number of papers~\cite{Babu:2015bna}-\cite{Saad:2022mzu}.
Our work differs from them in that we consider a more general model where the ${\bf 10}$ complex scalar field and its complex conjugate both have Yukawa couplings with {\bf 16} matter fields, as Peccei-Quinn charges~\cite{Peccei:1977hh} are not assigned to these fields.
As a result, the fitting of neutrino data becomes easier and we obtain an interesting finding that 
 the $SO(10)$ GUT model can be consistent with not only the normal hierarchy but also the inverted hierarchy of the neutrino mass
\footnote{
There is a recent report~\cite{Saad:2022mzu} that a model with a ${\bf 10}$ and a ${\bf 120}$ real scalar fields and a ${\bf 126}$ complex scalar field can also fit the inverted neutrino mass hierarchy.
}.
Another feature of our study is that we include information on the VEV of the {\bf 126} field in the fitting of neutrino data,
 which allows us to determine the portions of the SM Higgs field in the ${\bf 126}$ field ($c_3,c_4$ defined in Eq.~(\ref{higgsconstituents})).

This paper is organized as follows:
In Section~\ref{section-model}, we describe the non-supersymmetric renormalizable $SO(10)$ GUT model we consider.
In Section~\ref{section-gcu}, we analyze the gauge coupling unification conditions and evaluate the intermediate-scale VEV of the ${\bf 126}$ scalar field.
In Section~\ref{section-fitting}, we constrain the Yukawa couplings of the ${\bf 10}$ and ${\bf 126}$ scalar fields at the intermediate scale from experimental data on the quark and charged lepton masses and quark flavor mixings,
 express the neutrino mass matrix with these Yukawa couplings and the intermediate-scale VEV,
 and fit neutrino oscillation data to derive a prediction on neutrino observables.
In Section~\ref{section-check}, we inspect the validity of approximations made in Sections~\ref{section-gcu},\ref{section-fitting}.
Section~\ref{section-summary} summarizes the paper.
\\

\section{Non-supersymmetric Renormalizable $SO(10)$ GUT Model}
\label{section-model}

The model is a non-supersymmetric $SO(10)$ gauge theory with the following field content:
Three generations of left-handed Weyl spinors in {\bf 16} representation, denoted by ${\bf 16}^i$ with $i$ the flavor index;
a complex scalar field in {\bf 10}, denoted by ${\bf 10}_H$;
a complex scalar field in {\bf 126}, denoted by ${\bf 126}_H$;
a real scalar field in {\bf 54}, denoted by ${\bf 54}_H$.
The Yukawa couplings are given by
\bea
-{\cal L}_{\rm Yukawa}=(Y_{10})_{ij} \ {\bf 16}^i \ {\bf 10}_H \ {\bf 16}^j +  (Z_{10})_{ij} \ {\bf 16}^i \ {\bf 10}_H^* \ {\bf 16}^j + (Y_{126})_{ij} \ {\bf 16}^i \ {\bf 126}_H^* \ {\bf 16}^j + {\rm H.c.}
\eea

The gauge symmetry breaking proceeds as follows:
The component of ${\bf 54}_H$ with charge $({\bf 1},{\bf 1},{\bf 1})$ in the $SU(4)\times SU(2)_L\times SU(2)_R$ subgroup develops a VEV and breaks $SO(10)$.
We write the $SO(10)$-breaking scale as $\mu=\mu_{\rm GUT}$.
The effective theory below scale $\mu=\mu_{\rm GUT}$ is a Pati-Salam model with $SU(4)\times SU(2)_L\times SU(2)_R$ gauge group~\cite{Pati:1974yy}.
We assume that the $({\bf 1},{\bf 2},{\bf 2})$ component of ${\bf 10}_H$ and the $({\bf \ov{10}},{\bf 1},{\bf 3})+({\bf 10},{\bf 3},{\bf 1})+({\bf 15},{\bf 2},{\bf 2})$ components of ${\bf 126}_H$ 
 have mass much smaller than $\mu_{\rm GUT}$ and remain in the effective Pati-Salam model.
The $({\bf \ov{10}},{\bf 1},{\bf 3})$ and $({\bf 10},{\bf 3},{\bf 1})$ components have the same mass
 as a consequence of $D$-parity~\cite{Chang:1983fu,Chang:1984uy}.
We write the fields of these components as $({\bf 1},{\bf 2},{\bf 2})_H,({\bf \ov{10}},{\bf 1},{\bf 3})_H,({\bf 10},{\bf 3},{\bf 1})_H,({\bf 15},{\bf 2},{\bf 2})_H$, respectively, and
 write the fields of the $({\bf 4},{\bf 2},{\bf 1})+({\bf \ov{4}},{\bf 1},{\bf 2})$ components of ${\bf 16}^i$ fermions as $({\bf 4},{\bf 2},{\bf 1})^i+({\bf \ov{4}},{\bf 1},{\bf 2})^i$.
The effective Pati-Salam model contains the following Yukawa couplings:
\begin{align}
-{\cal L}_{\rm Yukawa}&\supset (Y_1)_{ij} \ ({\bf 4},{\bf 2},{\bf 1})^i ({\bf 1},{\bf 2},{\bf 2})_H ({\bf \ov{4}},{\bf 1},{\bf 2})^j 
\nn\\
&+(Z_1)_{ij} \ ({\bf 4},{\bf 2},{\bf 1})^i ({\bf 1},{\bf 2},{\bf 2})_H^* ({\bf \ov{4}},{\bf 1},{\bf 2})^j
\nn\\
&+(Y_{15})_{ij} \ ({\bf 4},{\bf 2},{\bf 1})^i ({\bf 15},{\bf 2},{\bf 2})_H^* ({\bf \ov{4}},{\bf 1},{\bf 2})^j
\nn\\
&+ \frac{1}{2}(Y_N)_{ij} \ ({\bf \ov{4}},{\bf 1},{\bf 2})^i ({\bf \ov{10}},{\bf 1},{\bf 3})_H^* ({\bf \ov{4}},{\bf 1},{\bf 2})^j 
+ \frac{1}{2}(Y_N)_{ij} \ ({\bf 4},{\bf 2},{\bf 1})^i ({\bf 10},{\bf 3},{\bf 1})_H^* ({\bf 4},{\bf 2},{\bf 1})^j
\nn\\
&+ {\rm H.c.},
\end{align}
 where the Yukawa couplings satisfy the following matching conditions at scale $\mu=\mu_{\rm GUT}$ at tree level~\cite{Aulakh:2002zr}:
\begin{align}
Y_1&= -2\sqrt{2} \, Y_{10},
\label{y1}\\
Z_1&= -2\sqrt{2} \, Z_{10},
\label{z1}\\
Y_{15}&= 8\sqrt{2} \, Y_{126},
\label{y15}\\
Y_N&= 8\, Y_{126}.
\label{yn}
\end{align}

Next, the component of the $({\bf \ov{10}},{\bf 1},{\bf 3})_H$ field with charge $({\bf 1},{\bf 1},0)$ in the $SU(3)_C\times SU(2)_L\times U(1)_Y$ subgroup develops a VEV, $v_{\rm PS}$, as
\bea
\langle ({\bf \ov{10}},{\bf 1},{\bf 3})_H \rangle = v_{\rm PS}
\eea
 and breaks $SU(4)\times SU(2)_R$.
We write the $SU(4)\times SU(2)_R$-breaking scale as $\mu=\mu_{\rm PS}$.
The effective theory below scale $\mu=\mu_{\rm PS}$ is the SM with $SU(3)_C\times SU(2)_L\times U(1)_Y$ gauge group.
The pair of $({\bf 1},{\bf 2},\pm\frac{1}{2})$ components in the $({\bf 1},{\bf 2},{\bf 2})_H$ field, denoted by $H_u,H_d$,
 and the pair of $({\bf 1},{\bf 2},\pm\frac{1}{2})$ components in the $({\bf 15},{\bf 2},{\bf 2})_H$ field, denoted by $\Phi_u,\Phi_d$,
 gain a mass matrix and the $H_u,\epsilon H_d^*,\Phi_u,\epsilon \Phi_d^*$ fields
 mix with each other ($\epsilon$ denotes the antisymmetric tensor in $SU(2)_L$ space).
This mass matrix is assumed to have one negative eigenvalue at the electroweak scale and three positive eigenvalues at ${\cal O}(v_{\rm PS}^2)$.
The eigenstate belonging to the negative eigenvalue is identified with the SM Higgs field, denoted by $H$.
We express the $H$ component of each field as
\begin{align}
&H_u = c_1 \, H + ...,
\nn\\
&\epsilon H_d^* = c_2 \, H + ....,
\nn\\
&\Phi_u = c_3 \, H + ...,
\nn\\
&\epsilon \Phi_d^* = c_4 \, H + ....,
\label{higgsconstituents}
\end{align}
 where $c_1,c_2,c_3,c_4$ are numbers satisfying $|c_1|^2+|c_2|^2+|c_3|^2+|c_4|^2=1$, and ``..." is an abbreviation for other mass eigenstates.
The $({\bf 4},{\bf 2},{\bf 1})^i$ and $({\bf \ov{4}},{\bf 1},{\bf 2})^i$ fermions are decomposed into the isospin-doublet quarks, isospin-singlet up-type quarks, isospin-singlet down-type quarks, isospin-doublet leptons, isospin-singlet charged leptons and singlet neutrinos, denoted by $q^i$, $u^{c\,i}$, $d^{c\,i}$, $\ell^i$, $e^{c\,i}$, $\nu^{c\,i}$, respectively.
This effective theory contains the following Yukawa couplings:
\bea
-{\cal L}_{\rm Yukawa}\supset (Y_u)_{ij} \ q^i \epsilon H u^{c\,j} + (Y_d)_{ij} \ q^i H^* d^{c\,j} + (Y_e)_{ij} \ \ell^i H^* e^{c\,j} + (Y_D)_{ij} \ \ell^i \epsilon H \nu^{c\,j}
+{\rm H.c.},
\eea
 where the Yukawa couplings satisfy the following matching conditions at scale $\mu=\mu_{\rm PS}$ at tree level:
\begin{align}
Y_u&= c_1 \, Y_1 - c_2 \, Z_1 - \frac{1}{2\sqrt{3}}c_4 \, Y_{15},
\label{yu}\\
Y_d&= -c_2^* \, Y_1 + c_1^* \, Z_1  + \frac{1}{2\sqrt{3}}c_3^* \, Y_{15},
\label{yd}\\
Y_e&= -c_2^* \, Y_1 + c_1^* \, Z_1 - \frac{\sqrt{3}}{2}c_3^* \, Y_{15},
\label{ye}\\
Y_D&= c_1 \, Y_1  - c_2 \, Z_1 + \frac{\sqrt{3}}{2}c_4 \, Y_{15}.
\label{ynu}
\end{align}
The singlet neutrinos gain Majorana mass term below,
\begin{align}
-{\cal L}_{\rm Yukawa}\supset \frac{1}{2\sqrt{2}}v_{\rm PS}(Y_N)_{ij} \, \nu^{c\,i}\nu^{c\,j}+{\rm H.c.},
\end{align}
 and are integrated out at scale $\mu=\mu_{\rm PS}$. Then the Weinberg operator is derived as
\begin{align}
-{\cal L}_{\rm eff}=\frac{1}{2}(C_\nu)_{ij}\, \ell^i \epsilon H \, \ell^j \epsilon H+{\rm H.c.},
\end{align}
 where $C_\nu$ satisfies at scale $\mu=\mu_{\rm PS}$ at tree level,
\begin{align}
C_\nu &= -\frac{\sqrt{2}}{v_{\rm PS}} Y_D Y_N^{-1} Y_D^T.
\label{cnu0}
\end{align}
\\

We comment that in the present model, the $H_u,\epsilon H_d^*$ fields and the $\Phi_u,\epsilon \Phi_d^*$ fields are allowed to have mixing terms
 because quartic terms in the $SO(10)$ gauge theory below,
\bea
\lambda \, {\bf 126}_H^* {\bf 126}_H^2 {\bf 10}_H + \lambda' \, {\bf 126}_H {\bf 126}_H^{*\,2} {\bf 10}_H + {\rm H.c.},
\eea
 give rise to quartic terms in the effective Pati-Salam model below,
\begin{align}
\tilde{\lambda} \, ({\bf \ov{10}},{\bf 1},{\bf 3})_H^*({\bf \ov{10}},{\bf 1},{\bf 3})_H({\bf 15},{\bf 2},{\bf 2})_H({\bf 1},{\bf 2},{\bf 2})_H 
+ \tilde{\lambda}' \, ({\bf \ov{10}},{\bf 1},{\bf 3})_H^*({\bf \ov{10}},{\bf 1},{\bf 3})_H({\bf 15},{\bf 2},{\bf 2})_H^*({\bf 1},{\bf 2},{\bf 2})_H  + {\rm H.c.}
\label{quartic-ps}
\end{align}
When the $({\bf 1},{\bf 1},0)$ component of the $({\bf \ov{10}},{\bf 1},{\bf 3})_H$ field develops the VEV $v_{\rm PS}$,
 Eq.~(\ref{quartic-ps}) yields mixing terms for the $H_u,\epsilon H_d^*$ fields and the $\Phi_u,\epsilon \Phi_d^*$ fields.
Remarkably, these mixing terms are ${\cal O}(v_{\rm PS}^2)$ and of the same order as the mass terms of the $({\bf 1},{\bf 2},{\bf 2})_H$ and $({\bf 15},{\bf 2},{\bf 2})_H$ fields.
As a result, large mixings of $H_u,\epsilon H_d^*$ and $\Phi_u,\epsilon \Phi_d^*$ can be realized.
\\

We comment on the strong CP problem and dark matter in the present model.
Unlike previous models of non-supersymmetric $SO(10)$ GUT, we do not assign $U(1)$ Peccei-Quinn charges~\cite{Peccei:1977hh} to the ${\bf 16}^i$ matter fields and the ${\bf 10}_H,{\bf 126}_H$ scalar fields.
Nevertheless, we can implement the Peccei-Quinn mechanism by introducing new ${\bf 16}$ and ${\bf \ov{16}}$ Weyl fermions and ${\bf 1}$ complex scalar
 and assigning $U(1)$ Peccei-Quinn charge $+1$ to ${\bf 16},{\bf \ov{16}}$ and $-2$ to ${\bf 1}$.
Then a KSVZ axion~\cite{Kim:1979if,Shifman:1979if} emerges and can solve the strong CP problem. It can also be a dark matter candidate.
\\

\section{Gauge Coupling Unification}
\label{section-gcu}

In this section and the next section, we adopt the following experimental values of the gauge coupling constants, quark and charged lepton masses and quark flavor mixings:
The QCD and QED gauge coupling constants in 5-quark-flavor QCD$\times$QED theory are fixed as $\alpha_s^{(5)}(M_Z)=0.1181$ and $\alpha^{(5)}(M_Z)=1/127.95$.
The lepton pole masses and $W$, $Z$, Higgs boson pole masses are taken from Particle Data Group~\cite{ParticleDataGroup:2022pth}.
We use the results of lattice calculations of the individual up and down quark masses, the strange quark mass, the charm quark mass and the bottom quark mass in $\ov{{\rm MS}}$ scheme reviewed in Ref.~\cite{FlavourLatticeAveragingGroupFLAG:2021npn}, 
 which read $m_u(2~{\rm GeV})=2.14(8)~{\rm MeV}$, $m_d(2~{\rm GeV})=4.70(5)~{\rm MeV}$~\cite{FermilabLattice:2018est,Giusti:2017dmp}, 
 $m_s(2~{\rm GeV})=93.40(57)~{\rm MeV}$~\cite{FermilabLattice:2018est,EuropeanTwistedMass:2014osg,Lytle:2018evc,Chakraborty:2014aca}, 
 $m_c(3~{\rm GeV})=0.988(11)~{\rm GeV}$~\cite{FermilabLattice:2018est,EuropeanTwistedMass:2014osg,Chakraborty:2014aca,Alexandrou:2014sha,Hatton:2020qhk}, 
 $m_b(m_b)=4.203(11)~{\rm GeV}$~\cite{FermilabLattice:2018est,Chakraborty:2014aca,Hatton:2021syc,Colquhoun:2014ica,ETM:2016nbo,Gambino:2017vkx}.
We use the top quark pole mass measured by CMS in Ref.~\cite{Sirunyan:2019zvx}, which reads $M_t=170.5(8)$~GeV.
We calculate the CKM mixing angles and CP phase from the Wolfenstein parameters in Ref.~\cite{ckmfitter}.
The above data are translated into the values of the quark and lepton Yukawa coupling matrices and the gauge coupling constants
 at scale $\mu=M_Z$ in $\ov{{\rm MS}}$ scheme
 with the help of the code~\cite{code} based on Refs.~\cite{Jegerlehner:2001fb}-\cite{threshold}.
\\

We analyze the gauge coupling unification conditions and evaluate the $SU(4)\times SU(2)_R$-breaking VEV $v_{\rm PS}$.
To this end, we solve the two-loop RG equations~\cite{Machacek:1983tz}-\cite{Machacek:1984zw} of SM, and match the theory with 
 the $SU(4)\times SU(2)_L\times SU(2)_R$ gauge theory containing 
 three generations of Weyl fermions $({\bf 4},{\bf 2},{\bf 1})^i,({\bf \ov{4}},{\bf 1},{\bf 2})^i$ and complex scalars 
 $({\bf 1},{\bf 2},{\bf 2})_H,({\bf \ov{10}},{\bf 1},{\bf 3})_H,({\bf 10},{\bf 3},{\bf 1})_H,({\bf 15},{\bf 2},{\bf 2})_H$ at scale $\mu=\mu_{\rm PS}$.
Then we calculate the two-loop RG equations of the $SU(4)\times SU(2)_L\times SU(2)_R$ gauge theory,
 and match the theory with the $SO(10)$ gauge theory at scale $\mu=\mu_{\rm GUT}$.
From the above matching conditions, we evaluate $v_{\rm PS}$.
Additionally, we evaluate the mass of the $SO(10)$ gauge boson that gains mass along the breaking of $SO(10)$ to $SU(4)\times SU(2)_L\times SU(2)_R$.

We make two approximations:
First, we approximate that the scalar particles decoupled at scale $\mu=\mu_{\rm PS}$ have a common mass $M_{H_{\rm PS}}$.
This has little impact on the evaluation of $v_{\rm PS}$ because the power of $M_{H_{\rm PS}}$ in the equation determining $v_{\rm PS}$ Eq.~(\ref{vpsequation}) is relatively small.
Second, when solving the two-loop RG equations of the gauge couplings of the $SU(4)\times SU(2)_L\times SU(2)_R$ gauge theory, we omit two-loop contributions involving the Yukawa couplings.
Later in Section~\ref{section-check} we will check that this approximation does not affect the result.

Given the approximation on the scalar particle masses, 
 the matching conditions around scale $\mu\sim\mu_{\rm PS}$ in $\ov{{\rm MS}}$ scheme are given by
\begin{align}
\frac{1}{g_{2R}^2(\mu)}-\frac{2}{48\pi^2} &= 
\frac{5}{3}\left(\frac{1}{g_1^2(\mu)}-\frac{1}{8\pi^2}\frac{28}{5}\ln\frac{M_{G({\bf 3},{\bf 1},\frac{2}{3})}}{\mu}
-\frac{1}{8\pi^2}\frac{21}{5}\ln\frac{M_{G({\bf 1},{\bf 1},1)}}{\mu} + \frac{1}{8\pi^2} \frac{67}{6} \ln\frac{M_{H_{\rm PS}}}{\mu} \right)
\nn\\
&- \frac{2}{3}\left(\frac{1}{g_3^2(\mu)}-\frac{1}{8\pi^2}\frac{7}{2}\ln\frac{M_{G({\bf 3},{\bf 1},\frac{2}{3})}}{\mu} + \frac{1}{8\pi^2} \frac{67}{6} \ln\frac{M_{H_{\rm PS}}}{\mu} -\frac{3}{48\pi^2} \right),
\label{psmatching1}\\
\frac{1}{g_{2L}^2(\mu)}-\frac{2}{48\pi^2} &= \frac{1}{g_{2}^2(\mu)} + \frac{1}{8\pi^2} \frac{71}{6} \ln\frac{M_{H_{\rm PS}}}{\mu}-\frac{2}{48\pi^2},
\label{psmatching2}\\
\frac{1}{g_{4}^2(\mu)}-\frac{4}{48\pi^2} &= \frac{1}{g_{3}^2(\mu)}-\frac{1}{8\pi^2}\frac{7}{2}\ln\frac{M_{G({\bf 3},{\bf 1},\frac{2}{3})}}{\mu} + \frac{1}{8\pi^2} \frac{67}{6} \ln\frac{M_{H_{\rm PS}}}{\mu} - \frac{3}{48\pi^2},
\label{psmatching3}
\end{align}
 and those around scale $\mu\sim\mu_{\rm GUT}$ in $\ov{{\rm MS}}$ scheme are given by
\begin{align}
\frac{1}{g_{10}^2(\mu)} -\frac{8}{48\pi^2} &= \frac{1}{g_{2R}^2(\mu)} - \frac{1}{8\pi^2}21\, \ln\frac{M_{G({\bf 6},{\bf 2},{\bf 2})}}{\mu} + \frac{1}{8\pi^2} \ln\frac{M_{H({\bf 1},{\bf 3},{\bf 3})}}{\mu} -\frac{2}{48\pi^2},
\label{so10matching1}\\
\frac{1}{g_{10}^2(\mu)} -\frac{8}{48\pi^2} &= \frac{1}{g_{2L}^2(\mu)} - \frac{1}{8\pi^2}21\, \ln\frac{M_{G({\bf 6},{\bf 2},{\bf 2})}}{\mu} + \frac{1}{8\pi^2} \ln\frac{M_{H({\bf 1},{\bf 3},{\bf 3})}}{\mu} -\frac{2}{48\pi^2},
\label{so10matching2}\\
\frac{1}{g_{10}^2(\mu)} -\frac{8}{48\pi^2} &= \frac{1}{g_4^2(\mu)} - \frac{1}{8\pi^2}14\, \ln\frac{M_{G({\bf 6},{\bf 2},{\bf 2})}}{\mu} + \frac{1}{8\pi^2}\frac{1}{3} \ln\frac{M_{H({\bf 6},{\bf 1},{\bf 1})}^2M_{H({\bf 20}',{\bf 1},{\bf 1})}^4}{\mu^6} -\frac{4}{48\pi^2},
\label{so10matching3}
\end{align}
 where $g_3,g_2,g_1$ denote the gauge couplings of the $SU(3)_C\times SU(2)_L\times U(1)_Y$ gauge theory ($g_1$ is in the GUT normalization),
 $g_4,g_{2L},g_{2R}$ denote those of the $SU(4)\times SU(2)_L\times SU(2)_R$ gauge theory, and $g_{10}$ denotes that of the $SO(10)$ gauge theory.
$M_{G({\bf 3},{\bf 1},\frac{2}{3})}$, $M_{G({\bf 1},{\bf 1},1)}$ denote the masses of the gauge bosons that become massive along the $SU(4)\times SU(2)_R$ 
 breaking (subscripts display the charges in $SU(3)_C\times SU(2)_L\times U(1)_Y$),
 and $M_{G({\bf 6},{\bf 2},{\bf 2})}$ denotes the mass of the gauge boson that becomes massive along the $SO(10)$ breaking into $SU(4)\times SU(2)_L\times SU(2)_R$.
$M_{H_{\rm PS}}$ denotes the common mass of the scalar particles decoupled at scale $\mu=\mu_{\rm PS}$,
 and $M_{H({\bf 6},{\bf 1},{\bf 1})}$, $M_{H({\bf 20}',{\bf 1},{\bf 1})}$, $M_{H({\bf 1},{\bf 3},{\bf 3})}$ denote the masses of the scalar particles decoupled at scale $\mu=\mu_{\rm GUT}$ (subscripts display the charges in $SU(4)\times SU(2)_L\times SU(2)_R$).
There are two complex scalar particles with the same charge $({\bf 6},{\bf 1},{\bf 1})$, and $M_{H({\bf 6},{\bf 1},{\bf 1})}$ should be regarded as the geometric mean of their masses.
The scalar particles with charge $({\bf 20}',{\bf 1},{\bf 1})$ and $({\bf 1},{\bf 3},{\bf 3})$ are real.

From Eqs.~(\ref{so10matching1}),(\ref{so10matching2}) and the particle content of the $SU(4)\times SU(2)_L\times SU(2)_R$ gauge theory,
 we see that $g_{2R}=g_{2L}$ holds at any scale, 
 even if the Yukawa couplings are not neglected when solving the RG equations of the gauge couplings of the $SU(4)\times SU(2)_L\times SU(2)_R$ gauge theory.
This is in accord with the fact that $D$-parity is unbroken when the VEV of ${\bf 54}_H$ breaks $SO(10)$.
From Eqs.~(\ref{psmatching1}),(\ref{psmatching2}) and the fact that $g_{2R}=g_{2L}$, we obtain the one-loop relation below,
\bea
\frac{M_{G({\bf 3},{\bf 1},\frac{2}{3})}^{21} M_{G({\bf 1},{\bf 1},1)}^{21}}{\mu^{42}} \frac{M_{H_{\rm PS}}^2}{\mu^2}
=\exp\left[8\pi^2\left(\frac{5}{g_1^2(\mu)}-\frac{3}{g_2^2(\mu)}-\frac{2}{g_3^2(\mu)}\right)+2\right].
\label{psmatching-final}
\eea
We solve the two-loop RG equations of SM and insert the result into Eq.~(\ref{psmatching-final}), thereby obtaining
\bea
M_{G({\bf 3},{\bf 1},\frac{2}{3})}^{21} M_{G({\bf 1},{\bf 1},1)}^{21} M_{H_{\rm PS}}^2 = e^2 ( 10^{13.70} \, {\rm GeV})^{44}.
\label{vpsequation}
\eea
From the above relation, we evaluate $v_{\rm PS}$. We note
 $M_{G({\bf 3},{\bf 1},\frac{2}{3})}^2 = g_4^2\,v_{\rm PS}^2$,
 $M_{G({\bf 1},{\bf 1},1)}^2 =  g_{2R}^2\,v_{\rm PS}^2$.
The value of $M_{H_{\rm PS}}$ has little impact on the evaluation of $v_{\rm PS}$ because of its comparably small power of 2.
If $M_{H_{\rm PS}}$ lies in a natural range of $0.3\,v_{\rm PS}>M_{H_{\rm PS}}>0.03\,v_{\rm PS}$, we get
\bea
v_{\rm PS}=10^{14.0} \, {\rm GeV}.
\label{vpsresult}
\eea

From Eqs.~(\ref{so10matching2}),(\ref{so10matching3}) and the fact that $g_{2R}=g_{2L}$, we obtain the one-loop relation below,
\bea
\frac{\mu}{M_{H({\bf 1},{\bf 3},{\bf 3})}}\frac{M_{H({\bf 6},{\bf 1},{\bf 1})}^{2/3} M_{H({\bf 20}',{\bf 1},{\bf 1})}^{4/3}}{\mu^2}\frac{M_{G({\bf 6},{\bf 2},{\bf 2})}^7}{\mu^7}
=\exp\left[8\pi^2\left(\frac{1}{g_{2L}^2(\mu)}-\frac{1}{g_4^2(\mu)}\right)+\frac{1}{3}\right].
\label{gutmatching-final}
\eea
We solve the two-loop RG equations of the $SU(4)\times SU(2)_L\times SU(2)_R$ gauge theory and insert the result into Eq.~(\ref{gutmatching-final}),
 thereby obtaining
\bea
\frac{M_{H({\bf 6},{\bf 1},{\bf 1})}^{2/3} M_{H({\bf 20}',{\bf 1},{\bf 1})}^{4/3}}{M_{H({\bf 1},{\bf 3},{\bf 3})}}M_{G({\bf 6},{\bf 2},{\bf 2})}^7
=e^{1/3}(10^{15.04}~{\rm GeV})^8.
\label{mgutequation}
\eea
If we assume a mild hierarchy among the scalar particle masses as
\bea
M_{H({\bf 6},{\bf 1},{\bf 1})} = M_{H({\bf 20}',{\bf 1},{\bf 1})} \simeq 10^{13.5}~{\rm GeV},
\\
M_{H({\bf 1},{\bf 3},{\bf 3})} \simeq 10^{16.5}~{\rm GeV},
\eea
 then we get $M_{G({\bf 6},{\bf 2},{\bf 2})}\simeq 6\times10^{15}$~GeV and the current bound on the $p\to e^+\pi^0$ partial lifetime
 as well as those of other nucleon decay modes are satisfied.
\\

\section{Fitting of Neutrino Data}
\label{section-fitting}

We constrain the Yukawa couplings of the $SU(4)\times SU(2)_L\times SU(2)_R$ gauge theory $Y_1,Z_1,Y_{15},Y_N$ at scale $\mu=\mu_{\rm PS}$,
 from experimental data on the quark and charged lepton masses and quark flavor mixings.
Then we express the neutrino mass matrix with $Y_1,Z_1,Y_{15},Y_N$ and $v_{\rm PS}$ in Eq.~(\ref{vpsresult}) based on the Type-1 seesaw mechanism.
Finally, we fit experimental data on the neutrino mixing angles and mass differences with $Y_1,Z_1,Y_{15},Y_N$ under the above constraints.

First, we calculate the up-type quark, down-type quark and charged lepton Yukawa couplings in SM at scale $\mu=\mu_{\rm PS}$ by solving the SM two-loop RG equations.
We take $\mu_{\rm PS}=10^{13.7} \, {\rm GeV}$, in accordance with Eq.~(\ref{vpsequation}).
The result is presented in Table~\ref{values} in the form of the singular values of the Yukawa coupling matrices and the parameters of the CKM matrix at scale $\mu=\mu_{\rm PS}$.
\begin{table}[H]
\begin{center}
  \caption{The singular values of the Yukawa coupling matrices and the CKM mixing angles and CP phase in SM
  at scale $\mu=\mu_{\rm PS}=10^{13.7}$~GeV.
  Also shown are the errors of the quark Yukawa couplings, propagated from the experimental errors of the corresponding masses,
 and maximal errors of the CKM parameters, obtained by assuming maximal correlation of experimental errors of the Wolfenstein parameters.}
   \begin{tabular}{|c||c|} \hline
                                      & Value  \\ \hline
    $y_u$           &2.98(11)$\times10^{-6}$          \\
    $y_c$           &0.001519(17)                               \\ 
    $y_t$            &0.4458(42)                                    \\ \hline
    $y_d$           &6.729(72)$\times10^{-6}$                     \\
    $y_s$           &0.00013369(82)                                 \\ 
    $y_b$           &0.006402(20)                                   \\ \hline
    $y_e$           &2.732$\times10^{-6}$                                 \\
    $y_\mu$           &0.0005767                               \\
    $y_\tau$           &0.009803                                  \\ \hline
    $\cos\theta_{13}^{\rm ckm}\sin\theta_{12}^{\rm ckm}$            & 0.22503(24)      \\
    $\cos\theta_{13}^{\rm ckm}\sin\theta_{23}^{\rm ckm}$           & 0.04576(77)         \\
    $\sin\theta_{13}^{\rm ckm}$                                                              & 0.00403(22)      \\
    $\delta_{\rm km}$~(rad)                                                                       &1.148(33)            \\ \hline
  \end{tabular}
  \label{values}
  \end{center}
\end{table}

\noindent
Due to $D$-parity, $Y_1,Z_1,Y_{15}$ at scale $\mu=\mu_{\rm PS}$ have symmetric flavor indices. 
Therefore, in the flavor basis where the isospin-doublet down-type quarks
 have a diagonal Yukawa coupling, 
 the up-type quark, down-type quark and charged lepton Yukawa coupling matrices $Y_u,Y_d,Y_e$ at scale $\mu=\mu_{\rm PS}$ can be written as
\begin{align}
Y_u &= V_{\rm CKM}^T    \begin{pmatrix} 
      y_u & 0 & 0 \\
      0 & y_c \, e^{2i\,d_2} & 0 \\
      0 & 0 & y_t \, e^{2i\,d_3} \\
   \end{pmatrix}V_{\rm CKM},
\label{yu-ydbasis}\\
Y_d&=   \begin{pmatrix} 
      y_d & 0 & 0 \\
      0 & y_s & 0 \\
      0 & 0 & y_b \\
   \end{pmatrix},
\label{yd-ydbasis}\\
Y_e&= U_e^T    \begin{pmatrix} 
      y_e & 0 & 0 \\
      0 & y_\mu & 0 \\
      0 & 0 & y_\tau \\
   \end{pmatrix}U_e,
\label{ye-ydbasis}
\end{align}
 where $y_u,y_c,y_t,y_d,y_s,y_b,y_e,y_\mu,y_\tau$ are given in Table~\ref{values},
 $V_{\rm CKM}$ is the CKM matrix whose parameters are given in Table~\ref{values}, $d_2,d_3$ are unknown phases and $U_e$ is an unknown unitary matrix.
$d_2,d_3,U_e$ are not constrained experimentally.
Using Eqs.~(\ref{yu})-(\ref{ye}), we can write $Y_1,Z_1,Y_{15}$ at scale $\mu=\mu_{\rm PS}$ as
\begin{align}
Y_1 &= \frac{1}{4c_3^*(|c_1|^2-|c_2|^2)}\left\{ 4c_1^* c_3^* Y_u + (c_1^*c_4+3c_2c_3^*)Y_d + (-c_1^*c_4+c_2c_3^*)Y_e \right\},
\label{y1-ydbasis}\\
Z_1&=  \frac{1}{4c_3^*(|c_1|^2-|c_2|^2)}\left\{ 4c_2^* c_3^* Y_u + (c_2^*c_4+3c_1c_3^*)Y_d + (-c_2^*c_4+c_1c_3^*)Y_e \right\},
\label{z1-ydbasis}\\
Y_{15}&= \frac{\sqrt{3}}{2}\frac{1}{c_3^*}(Y_d-Y_e)
\label{y15-ydbasis}
\end{align}
 with $Y_u,Y_d,Y_e$ given by Eqs.~(\ref{yu-ydbasis})-(\ref{ye-ydbasis}).
$Y_N$ is related to $Y_d-Y_e$ in the following way: 
Eqs.~(\ref{y15}),(\ref{yn}) give that $Y_N=Y_{15}/\sqrt{2}$ at scale $\mu=\mu_{\rm GUT}$.
Then $Y_N$ and $Y_{15}$ evolve from $\mu=\mu_{\rm GUT}$ to lower energy scales through different RG equations
  in the $SU(4)\times SU(2)_L\times SU(2)_R$ gauge theory.
At scale $\mu=\mu_{\rm PS}$, $Y_{15}$ is proportional to $Y_d-Y_e$ as Eq.~(\ref{y15-ydbasis}).
Hence, to relate $Y_N$ to $Y_d-Y_e$, we have to solve the RG equations of $Y_N$ and $Y_{15}$ in the $SU(4)\times SU(2)_L\times SU(2)_R$ gauge theory. 
Unfortunately, this is not possible because the RG equations depend on the Yukawa couplings $Y_1,Z_1,Y_{15}$ that are undetermined before the fitting analysis is finished.
Therefore, we approximate $Y_N=Y_{15}/\sqrt{2}$ at scale $\mu=\mu_{\rm PS}$.
Later in Section~\ref{section-check} we will assess the impact of this approximation after $Y_1,Z_1,Y_{15}$ are determined.
Given the above approximation, we can express the coefficient of the Weinberg operator at scale $\mu=\mu_{\rm PS}$
 using Eqs.~(\ref{y1-ydbasis})-(\ref{y15-ydbasis}) and Eqs.~(\ref{ynu}),(\ref{cnu0}) as
\begin{align}
C_\nu = \frac{2\sqrt{2}}{\sqrt{3}}c_3^*\frac{\sqrt{2}}{v_{\rm PS}} \left( Y_u + \frac{c_4}{c_3^*}(Y_d-Y_e) \right) (Y_d-Y_e)^{-1} \left( Y_u + \frac{c_4}{c_3^*}(Y_d-Y_e) \right)
\ \ \ \ \ 
{\rm at} \ \mu=\mu_{\rm PS}.
\label{cnu3}
\end{align}
\\

The fitting analysis is performed as follows:
First, we evaluate Eq.~(\ref{cnu3}) with the central values in Table~\ref{values}
 and the estimate of $v_{\rm PS}=10^{14.0}$~GeV in Eq.~(\ref{vpsresult}).
At this stage, phases $d_2,d_3$, unitary matrix $U_e$ and complex numbers $c_3,c_4$ are free parameters
 except that the latter two satisfy
\bea
|c_3|^2+|c_4|^2<1.
\eea
Next, we solve the 1-loop RG equation for the Wilson coefficient of the Weinberg operator $C_\nu$ from scale $\mu=\mu_{\rm PS}$ to $\mu=M_Z$,
 and evaluate the neutrino mass matrix as
\bea
M_\nu = \frac{v^2}{2}C_\nu(M_Z)
\eea
 with $v=246$~GeV.
From $M_\nu$ above, we derive the three neutrino mixing angles and the two neutrino mass differences.
Finally, we fit the neutrino oscillation data in NuFIT5.1 (with SK atmospheric data)~\cite{Esteban:2020cvm,nufit} with the free parameters $d_2,d_3,U_e,c_3,c_4$.
We consider {\bf both the normal hierarchy and the inverted hierarchy} of the neutrino mass.
We perform the fitting repeatedly and collect multiple fitting results in which
 two mixing angles $\sin^2\theta_{12},\sin^2\theta_{13}$ and the ratio of the neutrino mass differences $\Delta m^2_{21}/|\Delta m^2_{31}|$ are within the $2\sigma$ ranges and mixing angle $\sin^2\theta_{23}$ is within the $3\sigma$ range of the NuFIT5.1 data.
\\

\begin{figure}[H]
  \begin{center}
    \includegraphics[width=80mm]{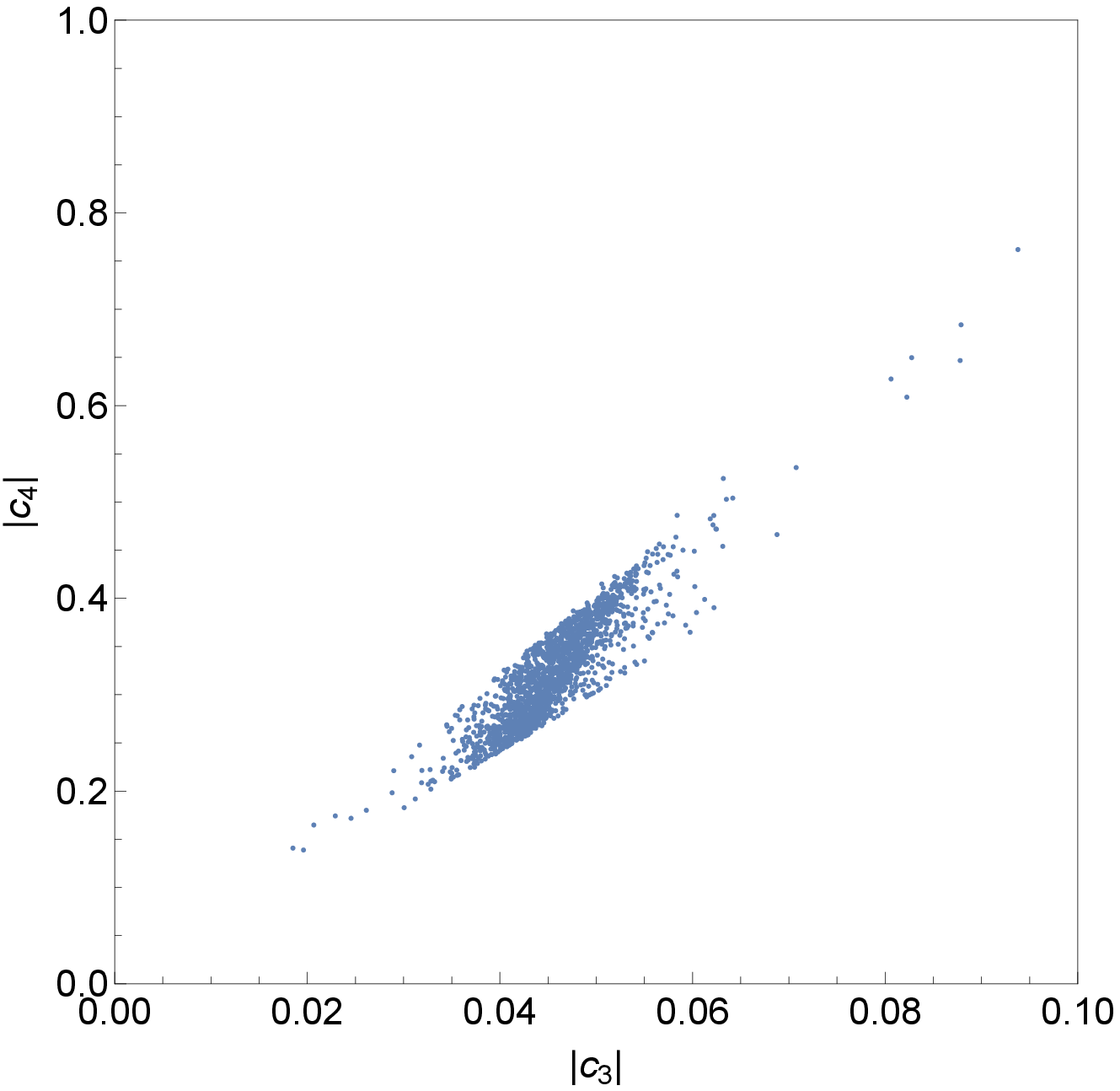}
    \includegraphics[width=80mm]{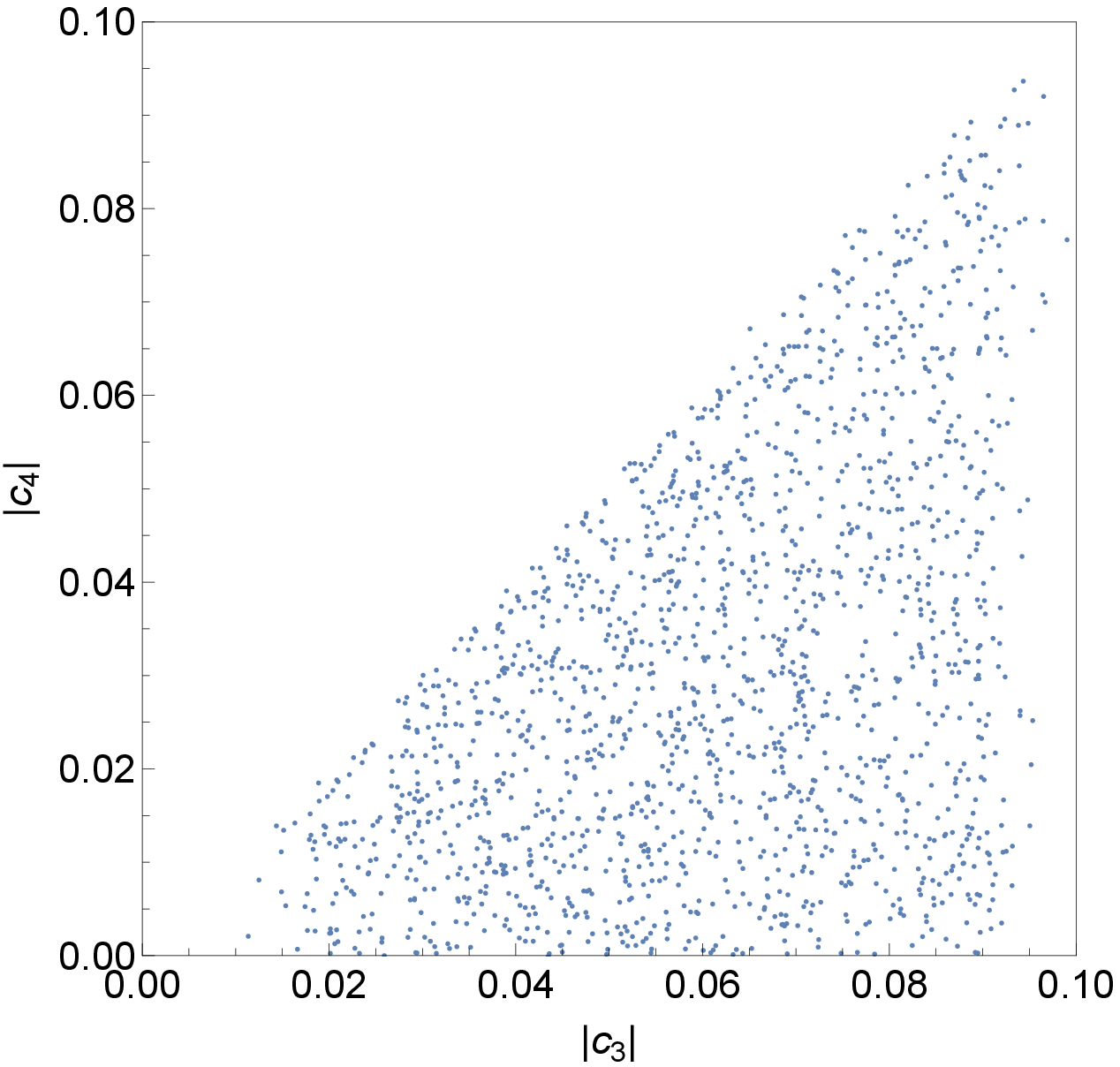}
    \caption{Fitting results on the plane of $|c_3|$ versus $|c_4|$, where $c_3,c_4$ quantify the portions of $({\bf 1},{\bf 2},\pm\frac{1}{2})$ components of $({\bf 15},{\bf 2},{\bf 2})_H$ field in the SM Higgs field as defined in Eq.~(\ref{higgsconstituents}).
 The left panel is for the case of the normal neutrino mass hierarchy and the right panel is for the case of the inverted hierarchy.}
    \label{c3c4}
  \end{center}
\end{figure}

We plot the fitting results on the plane of $|c_3|$ versus $|c_4|$ in Fig.~\ref{c3c4}.
Recall that $c_3,c_4$ quantify the portions of $({\bf 1},{\bf 2},\pm\frac{1}{2})$ components of $({\bf 15},{\bf 2},{\bf 2})_H$ field in the SM Higgs field as defined in Eq.~(\ref{higgsconstituents}).
The left panel is for the case of the normal neutrino mass hierarchy and the right panel is for the case of the inverted hierarchy.

\noindent
We see from Fig.~\ref{c3c4} that $|c_3|$ is ${\cal O}(0.01)$ in both normal and inverted hierarchy cases.
$|c_4|$ is ${\cal O}(10)$ times larger than $|c_3|$ in the normal hierarchy case, while it is on the same order or smaller than $|c_3|$ in the inverted hierarchy case. This implies that from the point of view of naturalness of the mass matrix of $({\bf 1},{\bf 2},\pm\frac{1}{2})$ components of $({\bf 15},{\bf 2},{\bf 2})_H$ field,
 the inverted hierarchy is favored because $|c_3|$ and $|c_4|$ can be on the same order.
Also, Fig.~\ref{c3c4} and Eq.~(\ref{cnu3}) indicate that the normal hierarchy is realized when the neutrino Dirac Yukawa coupling is dominated by the term proportional to $Y_d-Y_e$, whereas the inverted hierarchy is realized when the term proportional to $Y_u$ is dominant or comparable to that proportional to $Y_d-Y_e$.

\begin{figure}[H]
  \begin{center}
    \includegraphics[width=80mm]{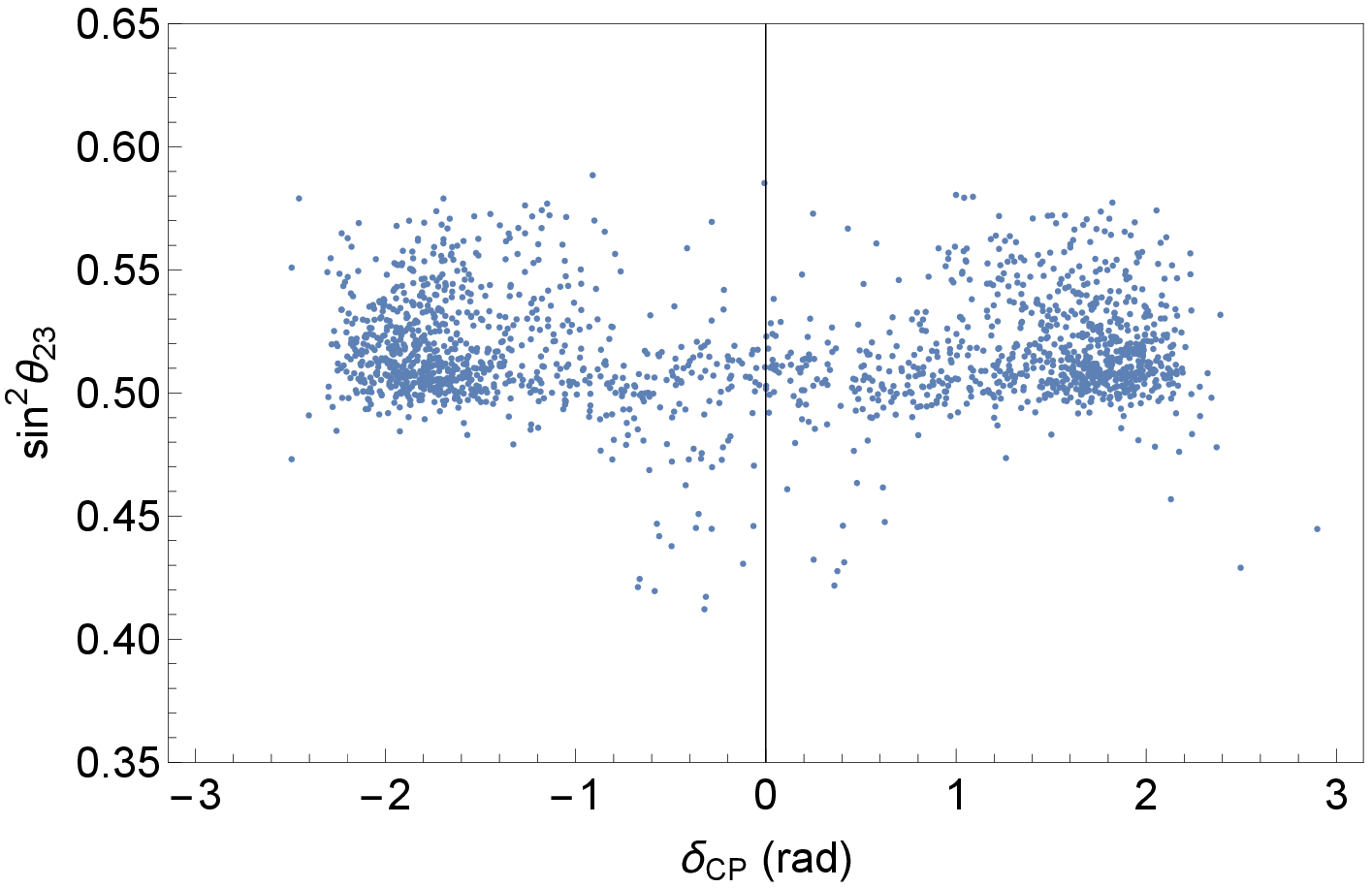}
    \includegraphics[width=80mm]{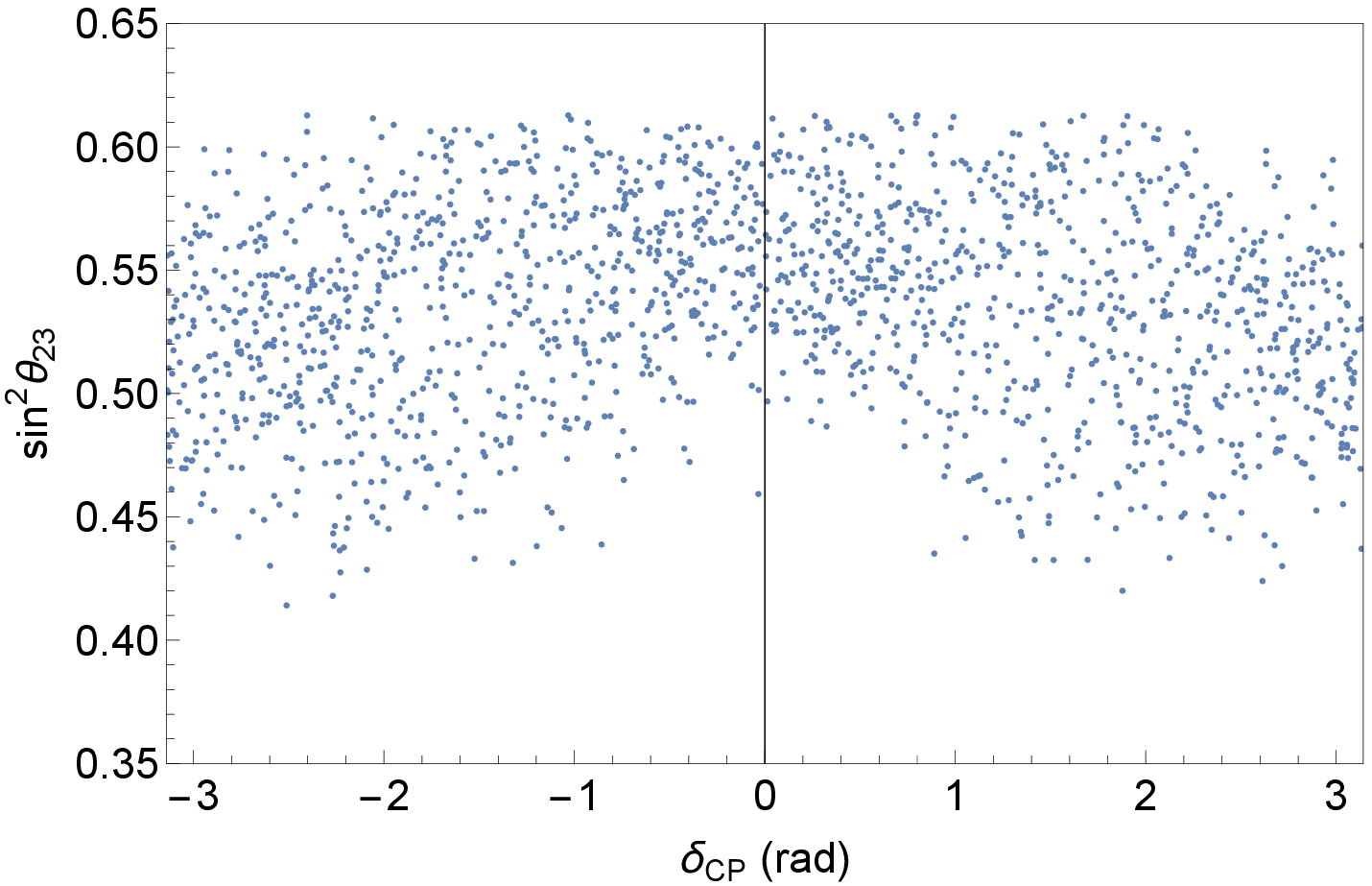}
    \\
    \includegraphics[width=80mm]{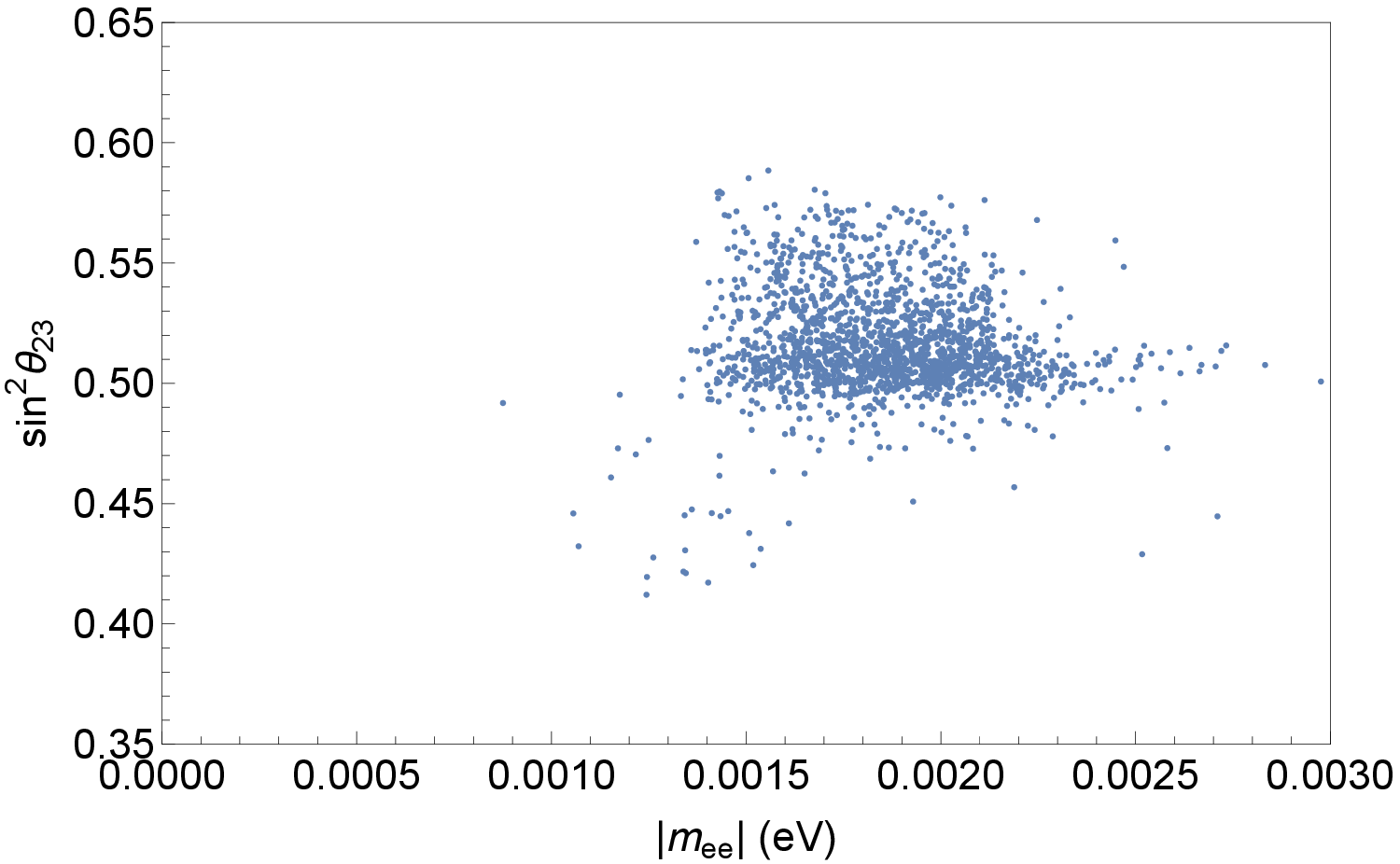}
    \includegraphics[width=80mm]{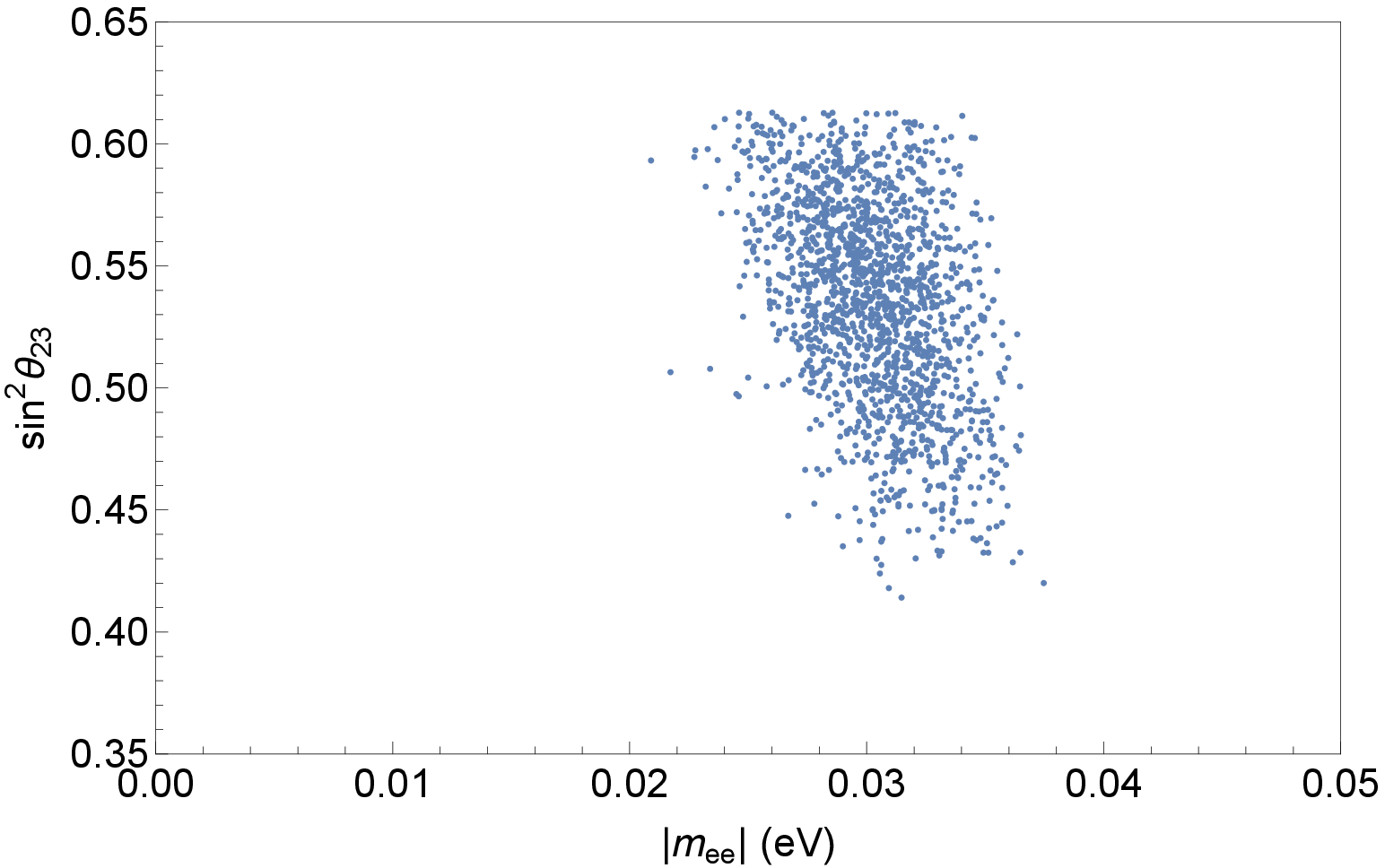}
    \\
    \includegraphics[width=80mm]{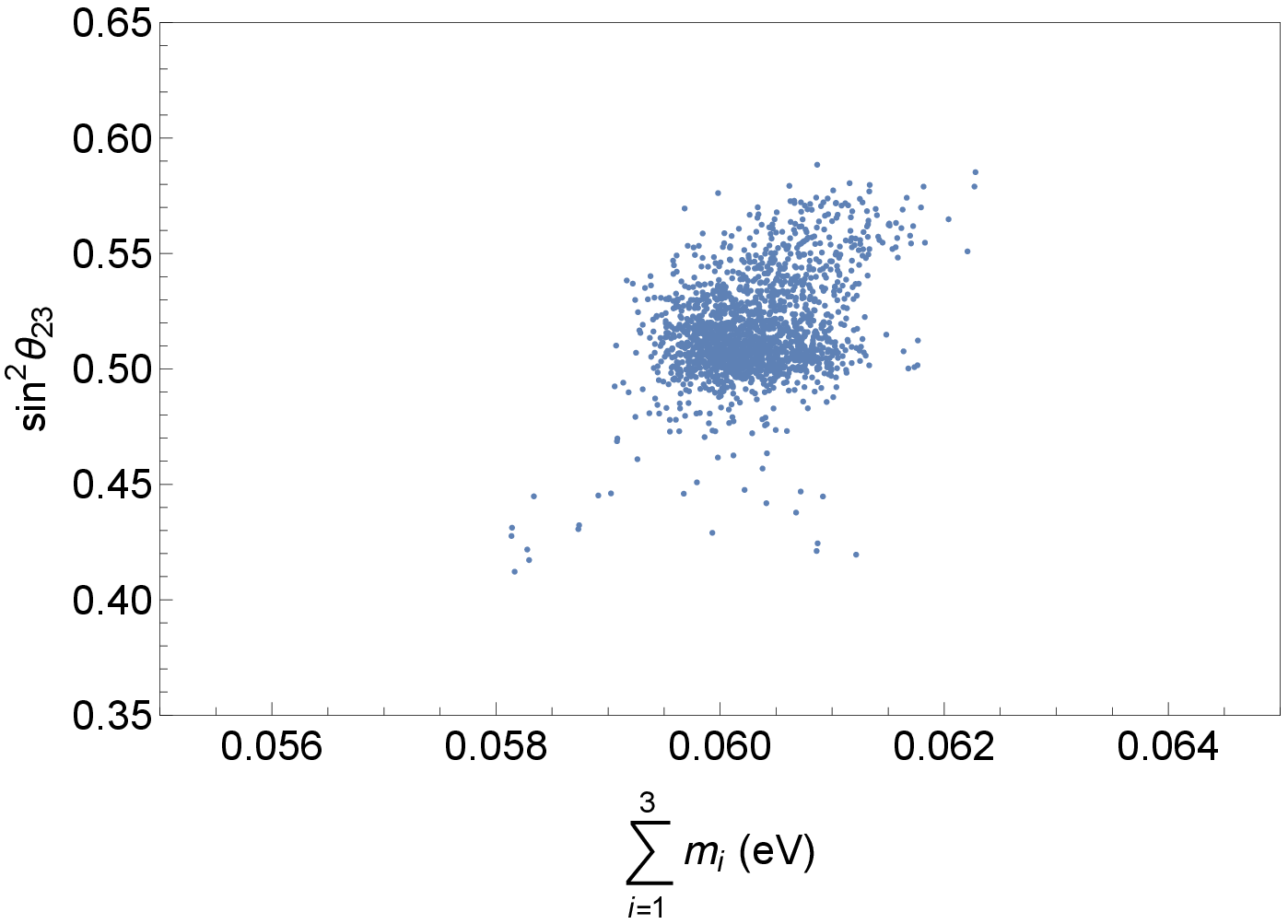}
    \includegraphics[width=80mm]{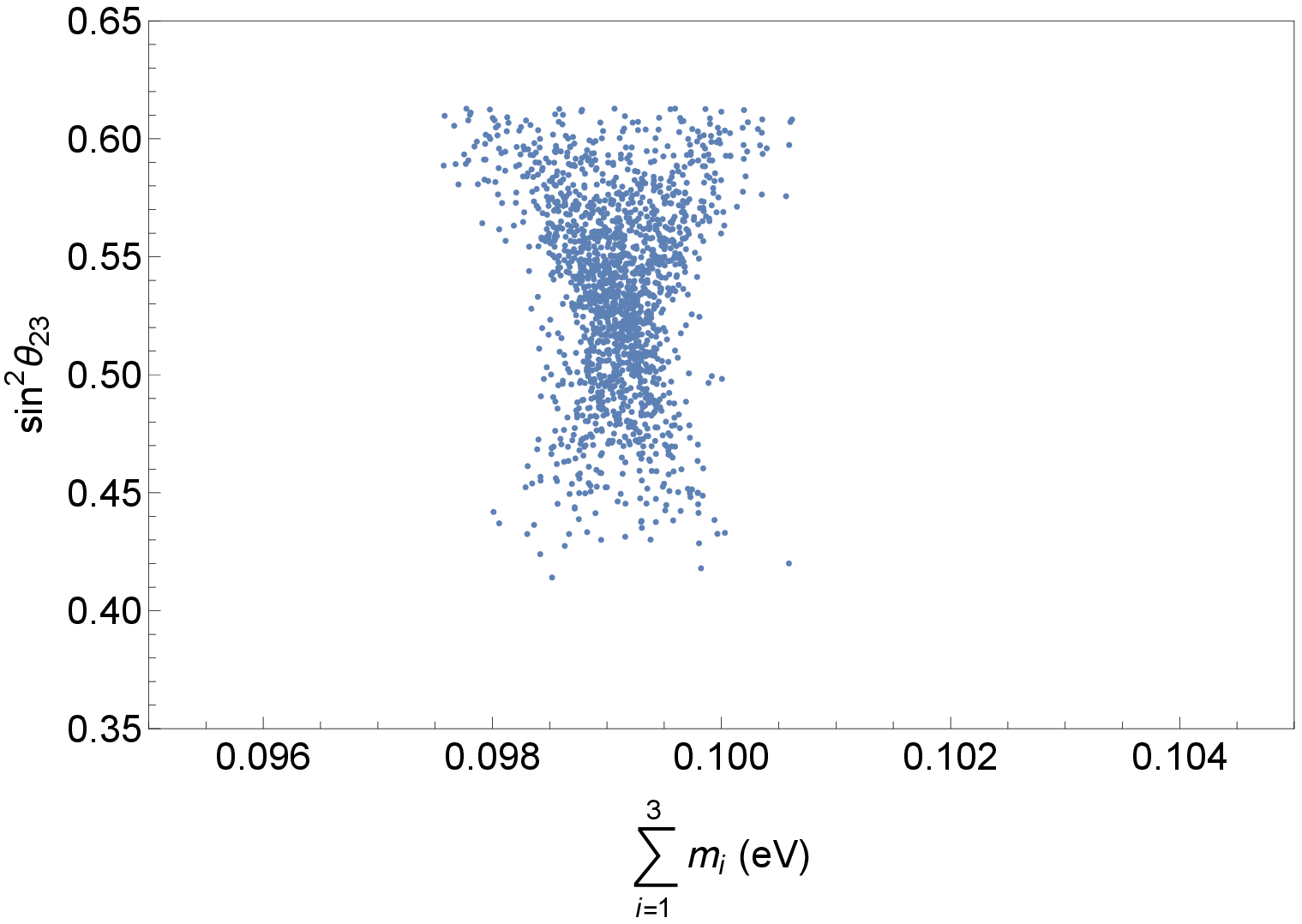}
    \caption{
    Panels in the first row: Fitting results on the plane of neutrino mixing angle $\sin^2\theta_{23}$ versus the Dirac CP phase of the neutrino mixing matrix $\delta_{\rm CP}$.
    Panels in the second row: Those on the plane of $\sin^2\theta_{23}$ versus the effective neutrino mass for neutrinoless double beta decay $|m_{ee}|$.
    Panels in the third row: Those on the plane of $\sin^2\theta_{23}$ versus the neutrino mass sum $\sum_{i=1}^3m_i$.
    The left-side panels are for the case of the normal neutrino mass hierarchy and the right-side panels are for the case of the inverted hierarchy.
    }
    \label{neutrinoobservables}
  \end{center}
\end{figure}

We examine the prediction of the model on neutrino observables, by plotting the fitting results on the planes of 
 neutrino mixing angle $\sin^2\theta_{23}$ versus the Dirac CP phase of the neutrino mixing matrix $\delta_{\rm CP}$,
 the effective neutrino mass for neutrinoless double beta decay $|m_{ee}|$, and
 the neutrino mass sum $\sum_{i=1}^3m_i$ in Fig.~\ref{neutrinoobservables}.
The left-side panels are for the case of the normal neutrino mass hierarchy and the right-side panels are for the case of the inverted hierarchy.
The reason that we focus on $\sin^2\theta_{23}$ is that it still has large uncertainty and further improvement of its measurement is anticipated.

\noindent
An interesting finding in Fig.~\ref{neutrinoobservables} is that in the normal hierarchy case, the Dirac CP phase $\delta_{\rm CP}$ is mostly in the ranges of
 $-2.4<\delta_{\rm CP}<-1.2$ and $1.2<\delta_{\rm CP}<2.4$, and it is unlikely that $\delta_{\rm CP}\sim\pi$.
This is in clear contrast with the inverted hierarchy case, where $\delta_{\rm CP}$ is equally distributed in the whole range.
Also, in the normal hierarchy case, mixing angle $\theta_{23}$ is likely in the upper octant and in a narrow range of $0.50\lesssim\sin^2\theta_{23}\lesssim0.55$.
The predictions on the effective neutrino mass for neutrinoless double beta decay $|m_{ee}|$ and the neutrino mass sum $\sum_{i=1}^3m_i$ are less interesting,
 since these predictions simply correspond to the situation with small lightest neutrino mass 
 ($m_1\ll \Delta m_{21}^2$ in the normal hierarchy case and $m_3\ll \Delta m_{21}^2$ in the inverted hierarchy case).

We further study the correlations between the predictions,
 by plotting the fitting results on the planes of $\delta_{\rm CP}$ versus $|m_{ee}|$; $\delta_{\rm CP}$ versus $\sum_{i=1}^3m_i$; $|m_{ee}|$ versus $\sum_{i=1}^3m_i$ in Fig.~\ref{neutrinoobservables2} in Appendix.
Unfortunately, we do not observe clear correlations among the predictions on $\delta_{\rm CP}$, $|m_{ee}|$ and $\sum_{i=1}^3m_i$.
\\

\section{Validity of the Approximations}
\label{section-check}

We have made two approximations: First, in Section~\ref{section-gcu}, we have neglected the contribution of the Yukawa couplings
 to the two-loop RG equations of the gauge couplings in the $SU(4)\times SU(2)_L\times SU(2)_R$ gauge theory.
Second, in Section~\ref{section-fitting}, we have neglected the RG evolutions of Yukawa couplings $Y_{15},Y_N$ in the $SU(4)\times SU(2)_L\times SU(2)_R$ gauge theory from scale $\mu=\mu_{\rm GUT}$ to $\mu=\mu_{\rm PS}$.

Now we inspect the validity of the above approximations.
For this purpose, we solve the full RG equations of the $SU(4)\times SU(2)_L\times SU(2)_R$ gauge theory including the Yukawa couplings $Y_1,Z_1,Y_{15},Y_N$.
Here the initial conditions of $Y_1,Z_1,Y_{15}$ at scale $\mu=\mu_{\rm PS}$ are given by Eqs.~(\ref{y1-ydbasis})-(\ref{y15-ydbasis}),(\ref{yu-ydbasis})-(\ref{ye-ydbasis})
 with the fitting results of Section~\ref{section-fitting} inserted into $d_2,d_3,U_e,c_3,c_4$.
Parameters $c_1,c_2$ have not been determined in the analysis of Section~\ref{section-fitting} and so we take $c_1=\sqrt{1-|c_3|^2-|c_4|^2}$ and $c_2=0$ to minimize the magnitudes of $Y_1,Z_1$.
We approximate $Y_N=Y_{15}/\sqrt{2}$ at scale $\mu=\mu_{\rm PS}$ 
 (which should hold exactly at scale $\mu=\mu_{\rm GUT}$, not at $\mu=\mu_{\rm PS}$)
 and study the relation between $Y_N$ and $Y_{15}/\sqrt{2}$ at scale $\mu=\mu_{\rm GUT}$ under the above approximation.

By solving the full RG equations of the $SU(4)\times SU(2)_L\times SU(2)_R$ gauge theory, we find that 
 the estimate of $10^{15.04}$~GeV in Eq.~(\ref{mgutequation}) is valid even if the two-loop contributions of the Yukawa couplings are included in the gauge coupling running.
We also find that the components of $Y_{15}/\sqrt{2}$ and $Y_N$ at scale $\mu=\mu_{\rm GUT}$ differ by at most 0.1\%,
 which implies that the approximation of taking $Y_N=Y_{15}/\sqrt{2}$ at scale $\mu=\mu_{\rm PS}$ (instead of at scale $\mu=\mu_{\rm GUT}$)
 has negligible impact compared to errors of the NuFIT5.1 data used in the fitting analysis.

\section{Summary}
\label{section-summary}

We have studied a prediction on neutrino observables in a non-supersymmetric renormalizable $SO(10)$ GUT model
 that contains a {\bf 10} complex scalar field and a {\bf 126} scalar field.
The {\bf 10} field and its complex conjugate and the complex conjugate of the {\bf 126} field have Yukawa couplings with the {\bf 16} matter fields $Y_{10},Z_{10},Y_{126}$,
 which give rise to the SM Yukawa couplings and neutrino Dirac Yukawa coupling.
The $SO(10)$ is broken into $SU(4)\times SU(2)_L\times SU(2)_R$ by the VEV of a {\bf 54} real scalar field, and it is broken into the SM gauge groups by the VEV of the {\bf 126} field.
The latter VEV and $Y_{126}$ generate Majorana mass for the singlet neutrinos.
For the above model, we have determined the VEV of the {\bf 126} field from the gauge coupling unification conditions.
We have constrained the Yukawa couplings of the ${\bf 10}$ and ${\bf 126}$ fields at the scale of the {\bf 126}  VEV
 from experimental data on the quark and charged lepton masses and quark flavor mixings,
 expressed the neutrino mass matrix with the above Yukawa couplings and the {\bf 126} VEV based on the Type-1 seesaw mechanism,
 fit the neutrino oscillation data, and derived a prediction on neutrino observables.

We have found that both the normal hierarchy and the inverted hierarchy of the neutrino mass can be fit
 and that the inverted hierarchy is favored from the point of view of naturalness of the mass matrix of $({\bf 1},{\bf 2},\pm\frac{1}{2})$ components of $({\bf 15},{\bf 2},{\bf 2})_H$ field.
Also, in the normal hierarchy case, the Dirac CP phase of the neutrino mixing matrix $\delta_{CP}$ is predicted to be likely in the ranges of 
 $-2.4<\delta_{\rm CP}<-1.2$ and $1.2<\delta_{\rm CP}<2.4$, and not in the range of $\delta_{\rm CP}\sim\pi$.
In the normal hierarchy case, mixing angle $\theta_{23}$ is predicted to be likely in the upper octant and in a narrow range of $0.50\lesssim\sin^2\theta_{23}\lesssim0.55$.

\section*{Acknowledgement}
This work is partially supported by Scientific Grants by the Ministry of Education, Culture, Sports, Science and Technology of Japan
 Nos.~21H000761 (NH) and No.~19K147101 (TY).
\\

\section*{Appendix}

\begin{figure}[H]
  \begin{center}    
    \includegraphics[width=80mm]{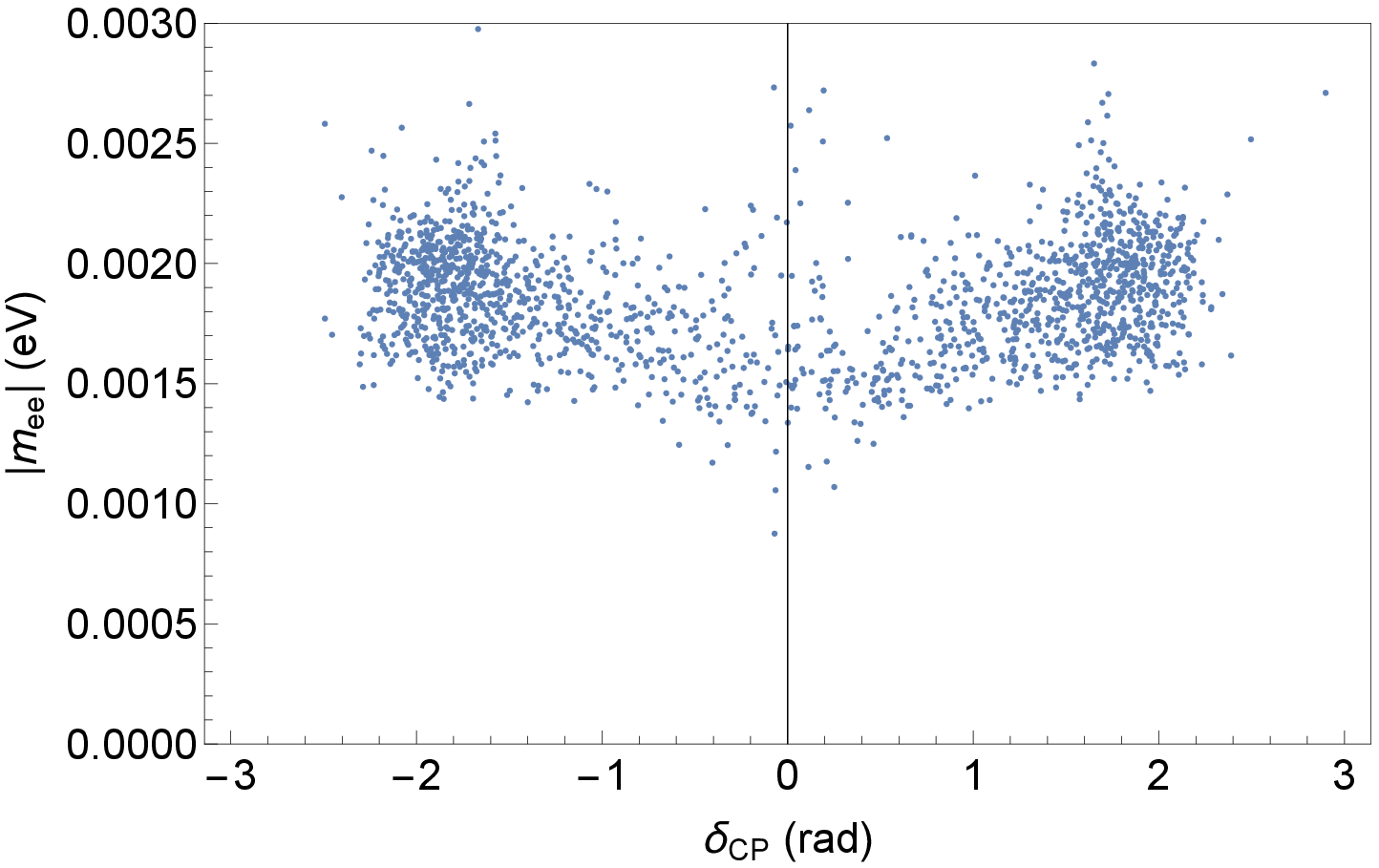}
    \includegraphics[width=80mm]{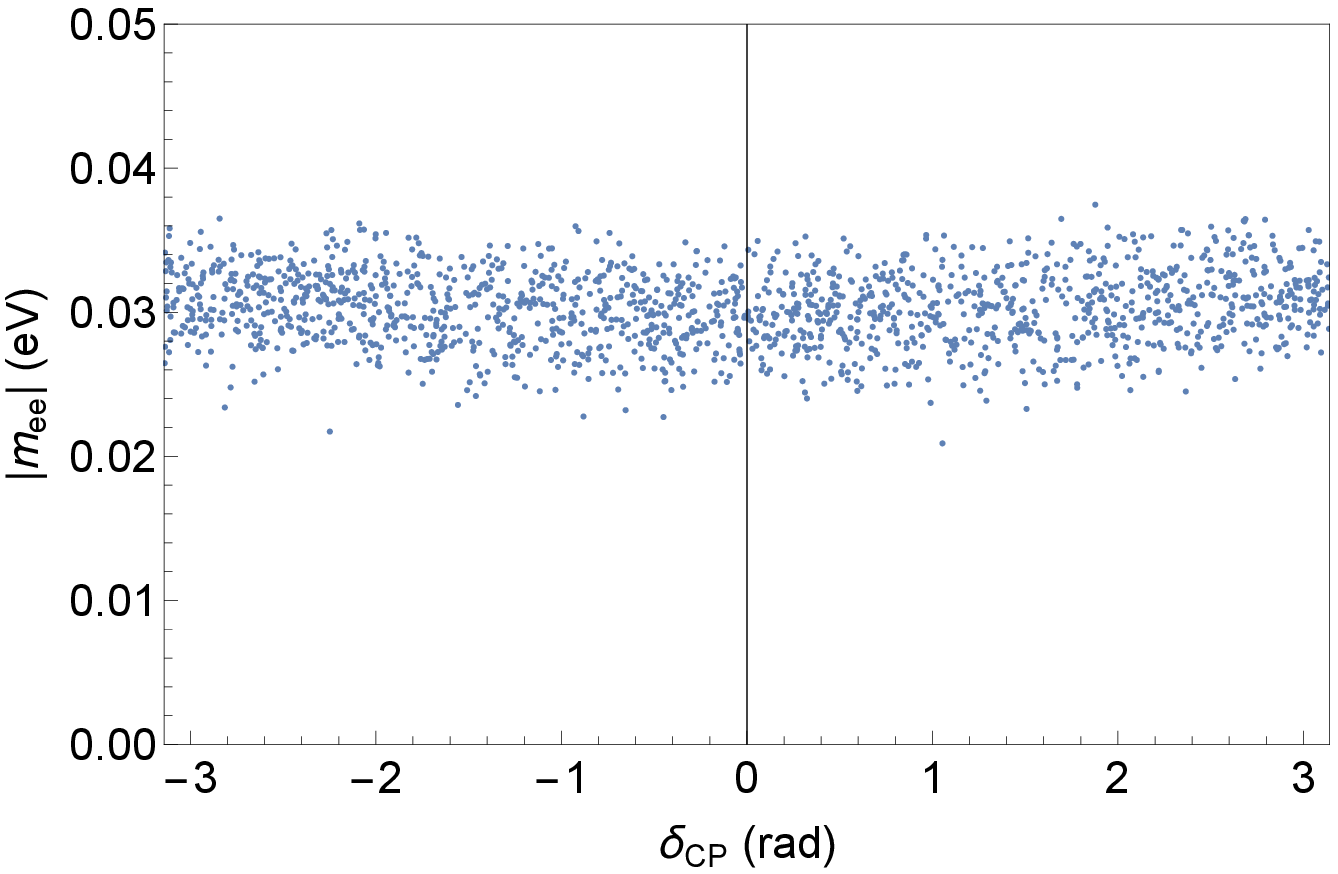}
    \\
    \includegraphics[width=80mm]{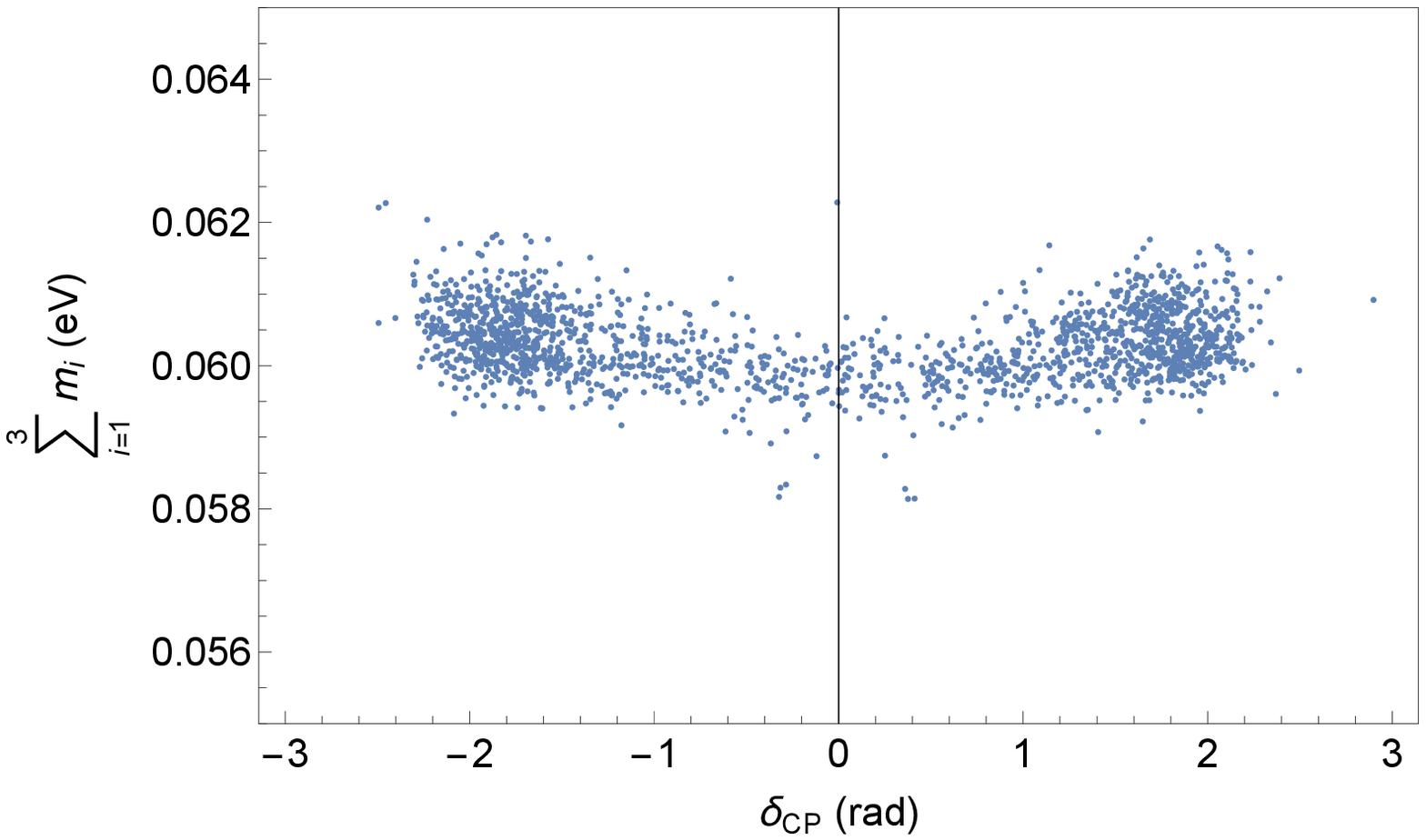}
    \includegraphics[width=80mm]{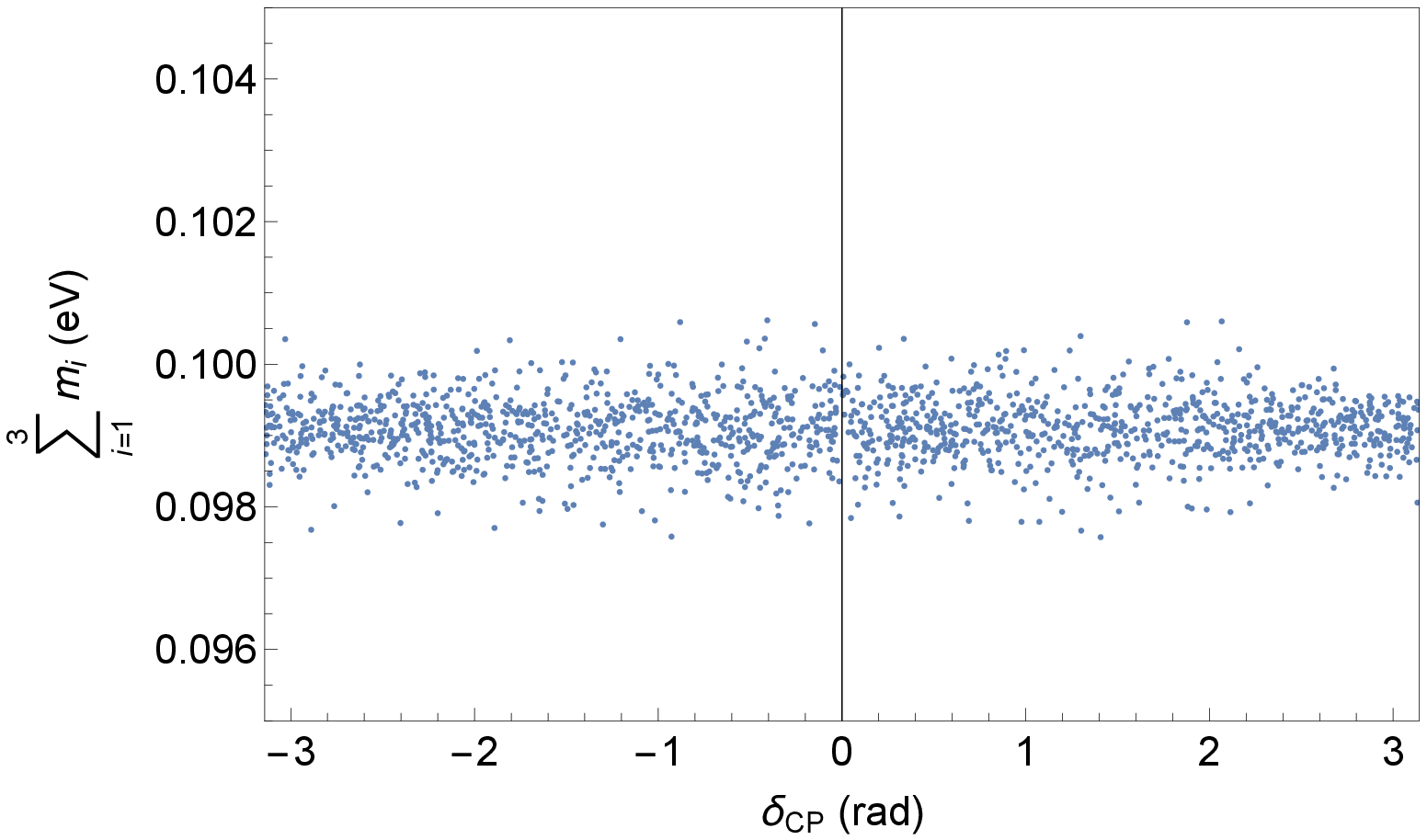}
    \\
    \includegraphics[width=80mm]{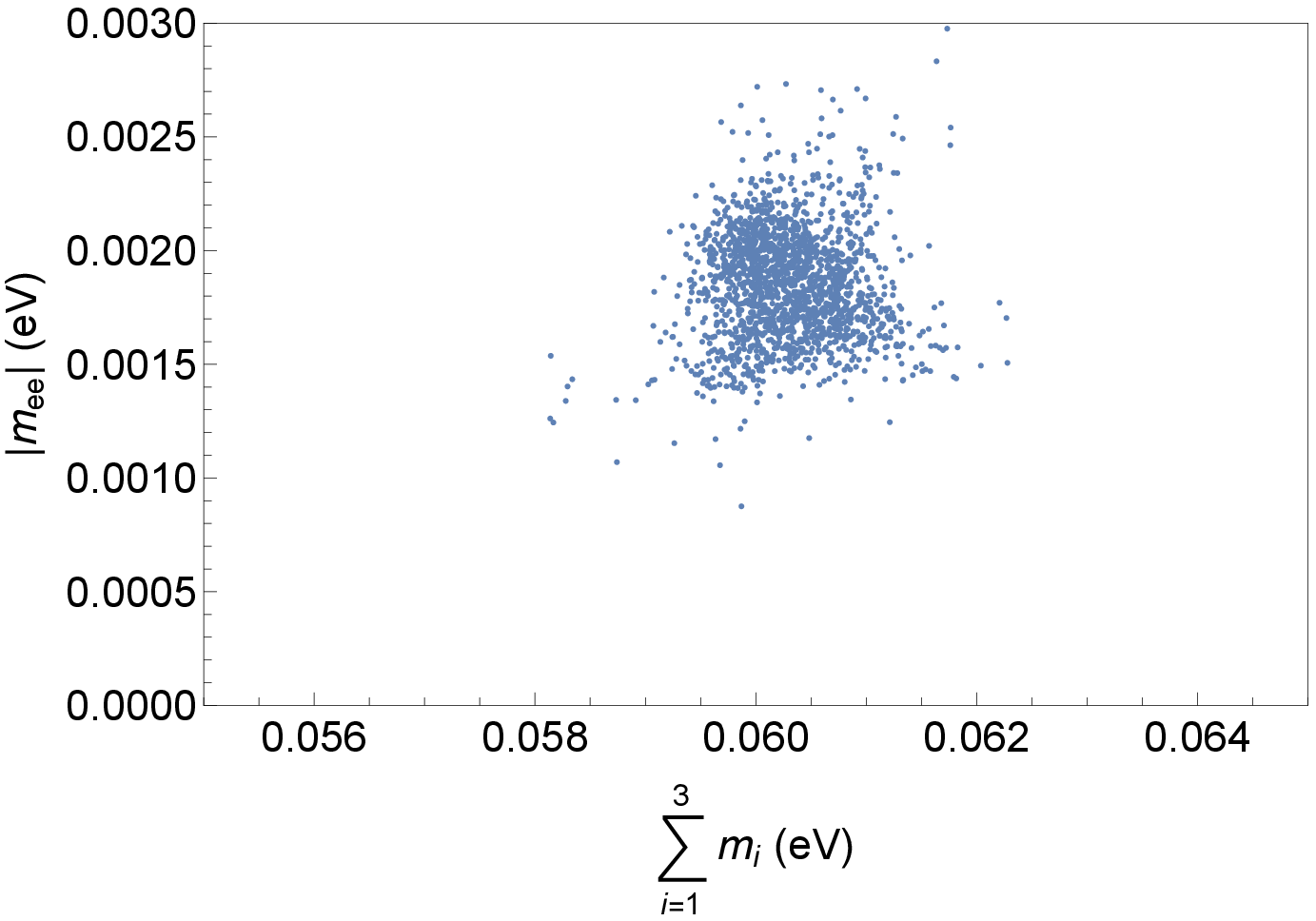}
    \includegraphics[width=80mm]{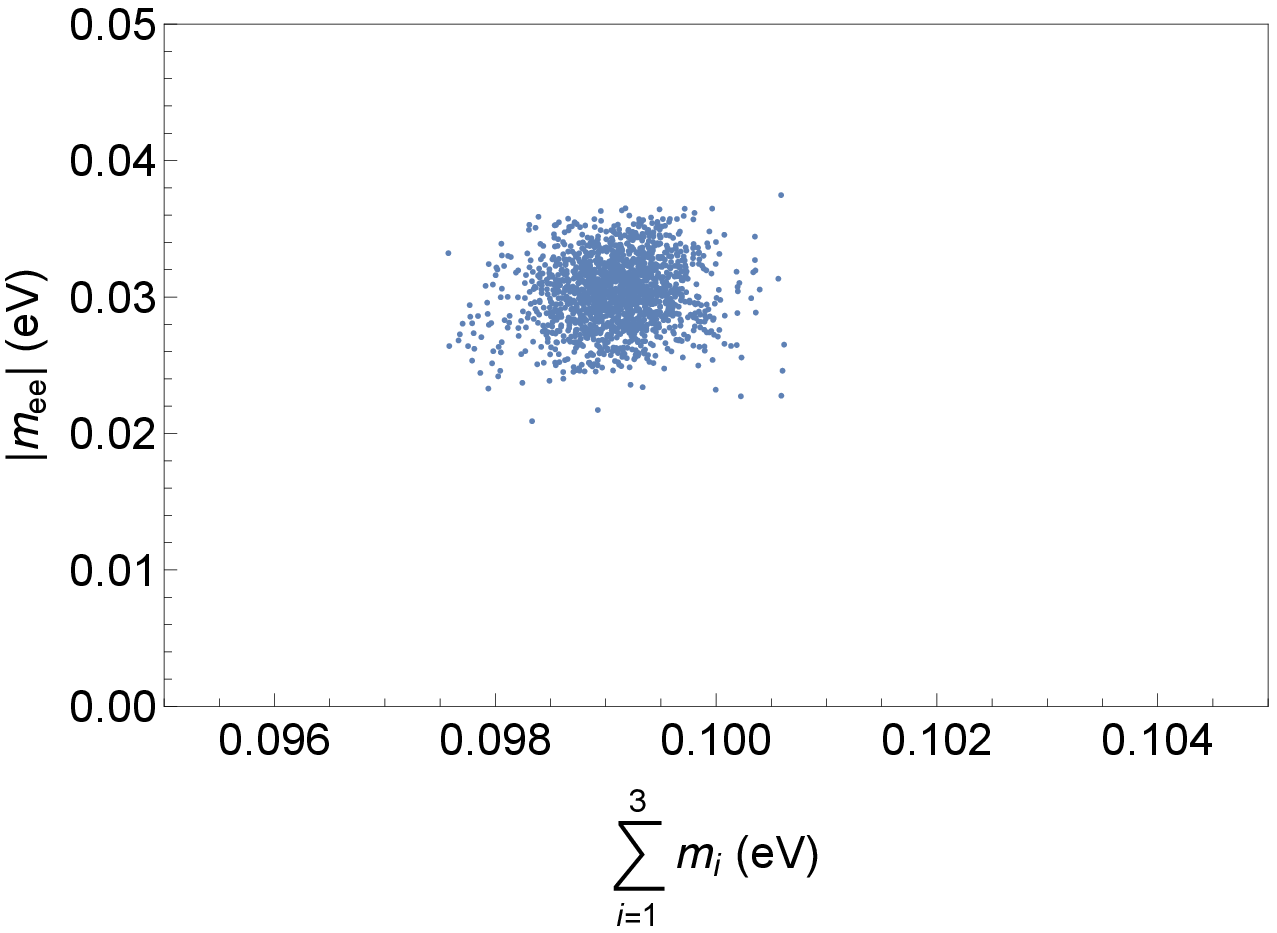}
    \caption{
    Panels in the first row: Fitting results on the plane of $\delta_{\rm CP}$ versus $|m_{ee}|$.
    Panels in the second row: Those on the plane of $\delta_{\rm CP}$ versus $\sum_{i=1}^3m_i$.
    Panels in the third row: Those on the plane of $|m_{ee}|$ versus $\sum_{i=1}^3m_i$.
    The left-side panels are for the case of the normal neutrino mass hierarchy, and the right-side panels are for the case of the inverted hierarchy.
    }
    \label{neutrinoobservables2}
  \end{center}
\end{figure}

To study the correlations between the predictions on neutrino observables, 
 we plot the fitting results of Section~\ref{section-fitting} on the planes of $\delta_{\rm CP}$ versus $|m_{ee}|$; $\delta_{\rm CP}$ versus $\sum_{i=1}^3m_i$; $|m_{ee}|$ versus $\sum_{i=1}^3m_i$
 in Fig.~\ref{neutrinoobservables2}.



\begin{thebibliography}{99}

\bibitem{Georgi:1974my} 
  H.~Georgi,
  ``The State of the Art—Gauge Theories,''
  AIP Conf.\ Proc.\  {\bf 23}, 575 (1975).
\bibitem{Fritzsch:1974nn} 
  H.~Fritzsch and P.~Minkowski,
  ``Unified Interactions of Leptons and Hadrons,''
  Annals Phys.\  {\bf 93}, 193 (1975).



\bibitem{Yanagida:1979as}
T.~Yanagida,
``Horizontal gauge symmetry and masses of neutrinos,''
Conf. Proc. C \textbf{7902131}, 95-99 (1979)
KEK-79-18-95.
\bibitem{Yanagida:1979gs}
T.~Yanagida,
``Horizontal Symmetry and Mass of the Top Quark,''
Phys. Rev. D \textbf{20}, 2986 (1979).
\bibitem{Gell-Mann:1979vob}
M.~Gell-Mann, P.~Ramond and R.~Slansky,
``Complex Spinors and Unified Theories,''
Conf. Proc. C \textbf{790927}, 315-321 (1979)
[arXiv:1306.4669 [hep-th]].
 \bibitem{seesaw4}
  R.~N.~Mohapatra and G.~Senjanovic,
  ``Neutrino Mass and Spontaneous Parity Violation,''
  Phys.\ Rev.\ Lett.\  {\bf 44}, 912 (1980).




\bibitem{Babu:1992ia}
K.~S.~Babu and R.~N.~Mohapatra,
``Predictive neutrino spectrum in minimal SO(10) grand unification,''
Phys. Rev. Lett. \textbf{70}, 2845-2848 (1993)
[arXiv:hep-ph/9209215 [hep-ph]].
\bibitem{Lee:1994je}
D.~G.~Lee and R.~N.~Mohapatra,
``Automatically R conserving supersymmetric SO(10) models and mixed light Higgs doublets,''
Phys. Rev. D \textbf{51}, 1353-1361 (1995)
[arXiv:hep-ph/9406328 [hep-ph]].




\bibitem{Babu:2015bna}
K.~S.~Babu and S.~Khan,
``Minimal nonsupersymmetric $SO(10)$ model: Gauge coupling unification, proton decay, and fermion masses,''
Phys. Rev. D \textbf{92}, no.7, 075018 (2015)
[arXiv:1507.06712 [hep-ph]].

\bibitem{Bajc:2005zf}
B.~Bajc, A.~Melfo, G.~Senjanovic and F.~Vissani,
``Yukawa sector in non-supersymmetric renormalizable SO(10),''
Phys. Rev. D \textbf{73}, 055001 (2006)
[arXiv:hep-ph/0510139 [hep-ph]].
\bibitem{Bertolini:2009qj}
S.~Bertolini, L.~Di Luzio and M.~Malinsky,
``Intermediate mass scales in the non-supersymmetric SO(10) grand unification: A Reappraisal,''
Phys. Rev. D \textbf{80}, 015013 (2009)
[arXiv:0903.4049 [hep-ph]].
\bibitem{Bertolini:2009es}
S.~Bertolini, L.~Di Luzio and M.~Malinsky,
``On the vacuum of the minimal nonsupersymmetric SO(10) unification,''
Phys. Rev. D \textbf{81}, 035015 (2010)
[arXiv:0912.1796 [hep-ph]].
\bibitem{Joshipura:2011nn}
A.~S.~Joshipura and K.~M.~Patel,
``Fermion Masses in SO(10) Models,''
Phys. Rev. D \textbf{83}, 095002 (2011)
[arXiv:1102.5148 [hep-ph]].
\bibitem{Bertolini:2012im}
S.~Bertolini, L.~Di Luzio and M.~Malinsky,
``Seesaw Scale in the Minimal Renormalizable SO(10) Grand Unification,''
Phys. Rev. D \textbf{85}, 095014 (2012)
[arXiv:1202.0807 [hep-ph]].
\bibitem{Buccella:2012kc}
F.~Buccella, D.~Falcone, C.~S.~Fong, E.~Nardi and G.~Ricciardi,
``Squeezing out predictions with leptogenesis from SO(10),''
Phys. Rev. D \textbf{86}, 035012 (2012)
[arXiv:1203.0829 [hep-ph]].
\bibitem{Awasthi:2013ff}
R.~L.~Awasthi, M.~K.~Parida and S.~Patra,
``Neutrino masses, dominant neutrinoless double beta decay, and observable lepton flavor violation in left-right models and SO(10) grand unification with low mass $ W_R, Z_R$ bosons,''
JHEP \textbf{08}, 122 (2013)
[arXiv:1302.0672 [hep-ph]].
\bibitem{Altarelli:2013aqa}
G.~Altarelli and D.~Meloni,
``A non supersymmetric SO(10) grand unified model for all the physics below $M_{GUT}$,''
JHEP \textbf{08}, 021 (2013)
[arXiv:1305.1001 [hep-ph]].
\bibitem{Dueck:2013gca}
A.~Dueck and W.~Rodejohann,
``Fits to SO(10) Grand Unified Models,''
JHEP \textbf{09}, 024 (2013)
[arXiv:1306.4468 [hep-ph]].
\bibitem{Kolesova:2014mfa}
H.~Kole\v{s}ov\'a and M.~Malinsk\'y,
``Proton lifetime in the minimal SO(10) GUT and its implications for the LHC,''
Phys. Rev. D \textbf{90}, no.11, 115001 (2014)
[arXiv:1409.4961 [hep-ph]].
\bibitem{Fong:2014gea}
C.~S.~Fong, D.~Meloni, A.~Meroni and E.~Nardi,
``Leptogenesis in SO(10),''
JHEP \textbf{01}, 111 (2015)
[arXiv:1412.4776 [hep-ph]].
\bibitem{Babu:2016bmy}
K.~S.~Babu, B.~Bajc and S.~Saad,
``Yukawa Sector of Minimal SO(10) Unification,''
JHEP \textbf{02}, 136 (2017)
[arXiv:1612.04329 [hep-ph]].
\bibitem{Schwichtenberg:2018cka}
J.~Schwichtenberg,
``Gauge Coupling Unification without Supersymmetry,''
Eur. Phys. J. C \textbf{79}, no.4, 351 (2019)
[arXiv:1808.10329 [hep-ph]].
\bibitem{Boucenna:2018wjc}
S.~M.~Boucenna, T.~Ohlsson and M.~Pernow,
``A minimal non-supersymmetric SO(10) model with Peccei--Quinn symmetry,''
Phys. Lett. B \textbf{792}, 251-257 (2019)
[erratum: Phys. Lett. B \textbf{797}, 134902 (2019)]
[arXiv:1812.10548 [hep-ph]].
\bibitem{Meloni:2019jcf}
D.~Meloni, T.~Ohlsson and M.~Pernow,
``Threshold effects in SO(10) models with one intermediate breaking scale,''
Eur. Phys. J. C \textbf{80}, no.9, 840 (2020)
[arXiv:1911.11411 [hep-ph]].
\bibitem{Hamada:2020isl}
Y.~Hamada, M.~Ibe, Y.~Muramatsu, K.~y.~Oda and N.~Yokozaki,
``Proton decay and axion dark matter in $SO$(10) grand unification via minimal left\textendash{}right symmetry,''
Eur. Phys. J. C \textbf{80}, no.5, 482 (2020)
[arXiv:2001.05235 [hep-ph]].
\bibitem{Ohlsson:2020rjc}
T.~Ohlsson, M.~Pernow and E.~S\"onnerlind,
``Realizing unification in two different SO(10) models with one intermediate breaking scale,''
Eur. Phys. J. C \textbf{80}, no.11, 1089 (2020)
[arXiv:2006.13936 [hep-ph]].
\bibitem{Fu:2022lrn}
B.~Fu, S.~F.~King, L.~Marsili, S.~Pascoli, J.~Turner and Y.~L.~Zhou,
``A predictive and testable unified theory of fermion masses, mixing and leptogenesis,''
JHEP \textbf{11}, 072 (2022)
[arXiv:2209.00021 [hep-ph]].
\bibitem{Lazarides:2022ezc}
G.~Lazarides, R.~Maji, R.~Roshan and Q.~Shafi,
``A predictive SO(10) model,''
JCAP \textbf{12}, 009 (2022)
[arXiv:2210.03710 [hep-ph]].
\bibitem{Saad:2022mzu}
S.~Saad,
``Probing Minimal Grand Unification through Gravitational Waves, Proton Decay, and Fermion Masses,''
[arXiv:2212.05291 [hep-ph]].






\bibitem{Peccei:1977hh}
R.~D.~Peccei and H.~R.~Quinn,
``CP Conservation in the Presence of Instantons,''
Phys. Rev. Lett. \textbf{38}, 1440-1443 (1977)








\bibitem{Pati:1974yy}
J.~C.~Pati and A.~Salam,
``Lepton Number as the Fourth Color,''
Phys. Rev. D \textbf{10}, 275-289 (1974)
[erratum: Phys. Rev. D \textbf{11}, 703-703 (1975)]




\bibitem{Chang:1983fu}
D.~Chang, R.~N.~Mohapatra and M.~K.~Parida,
``Decoupling Parity and SU(2)-R Breaking Scales: A New Approach to Left-Right Symmetric Models,''
Phys. Rev. Lett. \textbf{52}, 1072 (1984)
\bibitem{Chang:1984uy}
D.~Chang, R.~N.~Mohapatra and M.~K.~Parida,
``A New Approach to Left-Right Symmetry Breaking in Unified Gauge Theories,''
Phys. Rev. D \textbf{30}, 1052 (1984)





\bibitem{Weinberg:1981wj} 
  S.~Weinberg,
  ``Supersymmetry at Ordinary Energies. 1. Masses and Conservation Laws,''
  Phys.\ Rev.\ D {\bf 26}, 287 (1982).
\bibitem{Sakai:1981pk} 
  N.~Sakai and T.~Yanagida,
  ``Proton Decay in a Class of Supersymmetric Grand Unified Models,''
  Nucl.\ Phys.\ B {\bf 197}, 533 (1982).









\bibitem{Aulakh:2002zr}
C.~S.~Aulakh and A.~Girdhar,
Int. J. Mod. Phys. A \textbf{20}, 865-894 (2005)
doi:10.1142/S0217751X0502001X
[arXiv:hep-ph/0204097 [hep-ph]].








\bibitem{Kim:1979if}
J.~E.~Kim,
``Weak Interaction Singlet and Strong CP Invariance,''
Phys. Rev. Lett. \textbf{43}, 103 (1979)
\bibitem{Shifman:1979if}
M.~A.~Shifman, A.~I.~Vainshtein and V.~I.~Zakharov,
``Can Confinement Ensure Natural CP Invariance of Strong Interactions?,''
Nucl. Phys. B \textbf{166}, 493-506 (1980)












\bibitem{ParticleDataGroup:2022pth}
R.~L.~Workman \textit{et al.} [Particle Data Group],
``Review of Particle Physics,''
PTEP \textbf{2022}, 083C01 (2022)
   
   
   
   
   
   
\bibitem{FlavourLatticeAveragingGroupFLAG:2021npn}
Y.~Aoki \textit{et al.} [Flavour Lattice Averaging Group (FLAG)],
``FLAG Review 2021,''
Eur. Phys. J. C \textbf{82}, no.10, 869 (2022)
[arXiv:2111.09849 [hep-lat]].
\bibitem{FermilabLattice:2018est}
A.~Bazavov \textit{et al.} [Fermilab Lattice, MILC and TUMQCD],
``Up-, down-, strange-, charm-, and bottom-quark masses from four-flavor lattice QCD,''
Phys. Rev. D \textbf{98}, no.5, 054517 (2018)
[arXiv:1802.04248 [hep-lat]].
\bibitem{Giusti:2017dmp}
D.~Giusti, V.~Lubicz, C.~Tarantino, G.~Martinelli, F.~Sanfilippo, S.~Simula and N.~Tantalo,
``Leading isospin-breaking corrections to pion, kaon and charmed-meson masses with Twisted-Mass fermions,''
Phys. Rev. D \textbf{95}, no.11, 114504 (2017)
[arXiv:1704.06561 [hep-lat]].


\bibitem{EuropeanTwistedMass:2014osg}
N.~Carrasco \textit{et al.} [European Twisted Mass],
``Up, down, strange and charm quark masses with N$_f$ = 2+1+1 twisted mass lattice QCD,''
Nucl. Phys. B \textbf{887}, 19-68 (2014)
[arXiv:1403.4504 [hep-lat]].
\bibitem{Lytle:2018evc}
A.~T.~Lytle \textit{et al.} [HPQCD],
``Determination of quark masses from $\mathbf{n_f=4}$ lattice QCD and the RI-SMOM intermediate scheme,''
Phys. Rev. D \textbf{98}, no.1, 014513 (2018)
[arXiv:1805.06225 [hep-lat]].
\bibitem{Chakraborty:2014aca}
B.~Chakraborty, C.~T.~H.~Davies, B.~Galloway, P.~Knecht, J.~Koponen, G.~C.~Donald, R.~J.~Dowdall, G.~P.~Lepage and C.~McNeile,
``High-precision quark masses and QCD coupling from $n_f=4$ lattice QCD,''
Phys. Rev. D \textbf{91}, no.5, 054508 (2015)
[arXiv:1408.4169 [hep-lat]].


\bibitem{Alexandrou:2014sha}
C.~Alexandrou, V.~Drach, K.~Jansen, C.~Kallidonis and G.~Koutsou,
``Baryon spectrum with $N_f=2+1+1$ twisted mass fermions,''
Phys. Rev. D \textbf{90}, no.7, 074501 (2014)
[arXiv:1406.4310 [hep-lat]].
\bibitem{Hatton:2020qhk}
D.~Hatton \textit{et al.} [HPQCD],
``Charmonium properties from lattice $QCD$+QED : Hyperfine splitting, $J/\psi$ leptonic width, charm quark mass, and $a^c_\mu$,''
Phys. Rev. D \textbf{102}, no.5, 054511 (2020)
[arXiv:2005.01845 [hep-lat]].




\bibitem{Hatton:2021syc}
D.~Hatton, C.~T.~H.~Davies, J.~Koponen, G.~P.~Lepage and A.~T.~Lytle,
``Determination of $\overline{m}_b/\overline{m}_c$ and $\overline{m}_b$ from $n_f=4$ lattice QCD$+$QED,''
Phys. Rev. D \textbf{103}, no.11, 114508 (2021)
[arXiv:2102.09609 [hep-lat]].
\bibitem{Colquhoun:2014ica}
B.~Colquhoun, R.~J.~Dowdall, C.~T.~H.~Davies, K.~Hornbostel and G.~P.~Lepage,
``$\Upsilon$ and $\Upsilon^{\prime}$ Leptonic Widths, $a_{\mu}^b$ and $m_b$ from full lattice QCD,''
Phys. Rev. D \textbf{91}, no.7, 074514 (2015)
[arXiv:1408.5768 [hep-lat]].
\bibitem{ETM:2016nbo}
A.~Bussone \textit{et al.} [ETM],
``Mass of the b quark and B -meson decay constants from N$_f$=2+1+1 twisted-mass lattice QCD,''
Phys. Rev. D \textbf{93}, no.11, 114505 (2016)
[arXiv:1603.04306 [hep-lat]].
\bibitem{Gambino:2017vkx}
P.~Gambino, A.~Melis and S.~Simula,
``Extraction of heavy-quark-expansion parameters from unquenched lattice data on pseudoscalar and vector heavy-light meson masses,''
Phys. Rev. D \textbf{96}, no.1, 014511 (2017)
[arXiv:1704.06105 [hep-lat]].





\bibitem{Sirunyan:2019zvx}
A.~M.~Sirunyan \textit{et al.} [CMS],
``Measurement of $\mathrm{t\bar t}$ normalised multi-differential cross sections in pp collisions at $\sqrt s=13$ TeV, and simultaneous determination of the strong coupling strength, top quark pole mass, and parton distribution functions,''
Eur. Phys. J. C \textbf{80}, no.7, 658 (2020)
[arXiv:1904.05237 [hep-ex]].
\bibitem{ckmfitter}
  J.~Charles {\it et al.} [CKMfitter Group],
  ``CP violation and the CKM matrix: Assessing the impact of the asymmetric $B$ factories,''
  Eur.\ Phys.\ J.\ C {\bf 41}, no. 1, 1 (2005)
  [hep-ph/0406184],
  updated results and plots available at: http://ckmfitter.in2p3.fr
\bibitem{code}
     B.~A.~Kniehl, A.~F.~Pikelner and O.~L.~Veretin,
  ``mr: a C++ library for the matching and running of the Standard Model parameters,''
  Comput.\ Phys.\ Commun.\  {\bf 206}, 84 (2016)
  [arXiv:1601.08143 [hep-ph]]. 
  
  
  
 
\bibitem{Jegerlehner:2001fb} 
  F.~Jegerlehner, M.~Y.~Kalmykov and O.~Veretin,
  ``MS versus pole masses of gauge bosons: Electroweak bosonic two loop corrections,''
  Nucl.\ Phys.\ B {\bf 641}, 285 (2002)
  [hep-ph/0105304];
    F.~Jegerlehner, M.~Y.~Kalmykov and O.~Veretin,
  ``MS-bar versus pole masses of gauge bosons. 2. Two loop electroweak fermion corrections,''
  Nucl.\ Phys.\ B {\bf 658}, 49 (2003)
  [hep-ph/0212319].
\bibitem{Jegerlehner:2003py} 
  F.~Jegerlehner and M.~Y.~Kalmykov,
  ``O(alpha alpha(s)) correction to the pole mass of the t quark within the standard model,''
  Nucl.\ Phys.\ B {\bf 676}, 365 (2004)
  [hep-ph/0308216];
    F.~Jegerlehner and M.~Y.~Kalmykov,
  ``O(alpha alpha(s)) relation between pole- and MS-bar mass of the t quark,''
  Acta Phys.\ Polon.\ B {\bf 34}, 5335 (2003)
  [hep-ph/0310361].
\bibitem{Bezrukov:2012sa} 
  F.~Bezrukov, M.~Y.~Kalmykov, B.~A.~Kniehl and M.~Shaposhnikov,
  ``Higgs Boson Mass and New Physics,''
  JHEP {\bf 1210}, 140 (2012)
  [arXiv:1205.2893 [hep-ph]].
\bibitem{topthreshold}
     P.~Marquard, A.~V.~Smirnov, V.~A.~Smirnov and M.~Steinhauser,
  ``Quark Mass Relations to Four-Loop Order in Perturbative QCD,''
  Phys.\ Rev.\ Lett.\  {\bf 114}, no. 14, 142002 (2015)
  [arXiv:1502.01030 [hep-ph]].
   \bibitem{threshold}
   B.~A.~Kniehl, A.~F.~Pikelner and O.~L.~Veretin,
  ``Two-loop electroweak threshold corrections in the Standard Model,''
  Nucl.\ Phys.\ B {\bf 896}, 19 (2015)
  [arXiv:1503.02138 [hep-ph]]. 


  
   
   
\bibitem{Machacek:1983tz}
M.~E.~Machacek and M.~T.~Vaughn,
``Two Loop Renormalization Group Equations in a General Quantum Field Theory. 1. Wave Function Renormalization,''
Nucl. Phys. B \textbf{222}, 83-103 (1983)
\bibitem{Machacek:1983fi}
M.~E.~Machacek and M.~T.~Vaughn,
``Two Loop Renormalization Group Equations in a General Quantum Field Theory. 2. Yukawa Couplings,''
Nucl. Phys. B \textbf{236}, 221-232 (1984)
\bibitem{Machacek:1984zw}
M.~E.~Machacek and M.~T.~Vaughn,
``Two Loop Renormalization Group Equations in a General Quantum Field Theory. 3. Scalar Quartic Couplings,''
Nucl. Phys. B \textbf{249}, 70-92 (1985)


\bibitem{Esteban:2020cvm}
I.~Esteban, M.~C.~Gonzalez-Garcia, M.~Maltoni, T.~Schwetz and A.~Zhou,
``The fate of hints: updated global analysis of three-flavor neutrino oscillations,''
JHEP \textbf{09}, 178 (2020)
[arXiv:2007.14792 [hep-ph]].
 \bibitem{nufit}
  NuFIT 5.1 (2021), www.nu-fit.org.



 
\end{thebibliography}
\end{document}